\documentclass[structabstract]{aa}  
\usepackage{graphicx}
\usepackage{txfonts}
\usepackage{natbib}
\usepackage{longtable,lscape}
\begin{document}
   \title{Revealing the ``missing'' low-mass stars in the S254-S258
star forming region by deep X-ray imaging\thanks{Tables 2, 3, and 5 are     
only available in electronic form
at the CDS via anonymous ftp to cdsarc.u-strasbg.fr (130.79.128.5)
or via http://cdsweb.u-strasbg.fr/cgi-bin/qcat?J/A+A/}}

   \author{P.~Mucciarelli
          \inst{1,2}
          \and
          T.~Preibisch\inst{1}
         \and
          H.~Zinnecker \inst{3,4,5}
          }

   \institute{Universit\"ats-Sternwarte M\"unchen,  Ludwig-Maximilians-Universit\"at,
               Scheinerstr.~1, 81679 M\"unchen, Germany\\
              \email{pmuccia@usm.uni-muenchen.de,preibisch@usm.uni-muenchen.de}
         \and
             Exzellenzcluster Universe, Boltzmannstr.~2, 85748 Garching, Germany
          \and   
	  	Astrophysikalisches Institut Potsdam, An der Sternwarte 16, 14482 Potsdam, Germany
\and
   Deutsches SOFIA Institut, Universit\"at Stuttgart, Pfaffenwaldring 31,
    70569 Stuttgart, Germany
    \and
NASA-Ames Research Center,
MS 211-3, Moffett Field, CA 94035, USA
		}

\titlerunning{Revealing the ``missing'' low-mass stars in the S254-S258
star forming region}
\authorrunning{Mucciarelli, Preibisch, \& Zinnecker}

\date{Received 13 April 2011; accepted 4 June 2011}

 
  \abstract
   {X-ray observations provide a very good way to reveal the 
population of young stars in star forming regions avoiding the biases 
introduced when selecting samples based on infrared excess.}
   {The aim of this study was to find an explanation for the
remarkable morphology of the central part of the S254--S258 star 
forming complex, where
a dense embedded cluster of very young stellar objects (S255-IR) 
is sandwiched between the
two \ion{H}{II} regions S255 and S257. 
This interesting configuration had led to 
different speculations such as dynamical ejection of the B-stars from the central cluster
or triggered star formation in a cloud that was swept up in the collision zone between
the two expanding \ion{H}{II} regions.
The presence or absence, and the spatial distribution of low-mass stars associated with these
B-stars can discriminate between the
possible scenarios.}
   {We performed a deep \textit{Chandra} X-ray observation of the S254--S258
region in order to 
efficiently discriminate young stars (with and without circumstellar matter)
from the numerous
older field stars in the area.
}
   {We detected 364 X-ray point sources in a $17' \times 17'$ field ($\approx 8 \times 8$~pc).
This X-ray catalog  provides, for the first time, a complete 
sample of all young stars in the region down 
to $\sim 0.5\,M_\odot$. 
A clustering analysis identifies three significant clusters: the
central embedded cluster S255-IR and two smaller clusterings in S256 and S258.
Sixty-four X-ray sources can be classified
as members in one of these clusters. 
After accounting for X-ray background contaminants, this implies that about 250 X-ray sources
constitute a widely scattered population of young stars, distributed over the full field-of-view
of our X-ray image. 
This distributed young stellar population is considerably larger than the previously 
known number of non-clustered young stars selected by infrared excesses.
Comparison of the X-ray luminosity function 
with that of the Orion Nebula Cluster suggests
a total population of $\sim 2000$ young stars in the observed part of the S254-S258 region.
}
   { The observed number of $\sim 250$ X-ray detected distributed young stars agrees 
well with the expectation for the low-mass population associated
to the B-stars in S255 and S257 as predicted by an IMF extrapolation.
These results are consistent with  the scenario
that these two B-stars represent an earlier stellar population
and that their expanding \ion{H}{II} regions have swept up the central cloud 
and trigger star formation (i.e.~the central embedded cluster S255-IR) therein.
}


   \maketitle
%

\section{Introduction}

\subsection{The S254-S258 complex}

The south-eastern part of the molecular cloud complex in the Gem OB1 association contains an 
embedded star-forming region with several diffuse \ion{H}{II} regions 
(S254--S258, \citealt{sha59}, see Fig.~1). 
The most prominent of these \ion{H}{II} regions, S255 and S257 \citep{Chopinet74}, are both powered by
B0 stars, have diameters of $\sim 4'$, and a projected separation of  $\sim 6'$. 
Sandwiched right between them is
a dense dusty molecular cloud filament {}(S255-IR or S255-2; \citealt{hey89,dif08}) that
contains numerous embedded infrared sources
\citep{zin93,how97,ito01,lon06,ojh06,cha08}. Masers, HH-objects, jets, and molecular outflows
\citep{sne86,mir97,min05,min07,god07,jia08,wan11} provide clear evidence of very recent and 
ongoing star formation activity in this cloud.
The combination of {\em Spitzer} mid-infrared observations
 \citep{all05} with near-infrared
images of a $26' \times 20'$ region 
led to the detection of 510 sources with near- or mid-IR excess \citep{cha08},
87 and 165 of which were classified as Class~I and Class~II sources, respectively.
The large majority (80\%) of these infrared excess-selected young stellar objects (YSOs)
 were found to be clustered.
The central cluster S255-IR is the richest of these.
It contains at least\footnote{The census is incomplete because the
{\em Spitzer} images of the dense cluster suffer from source crowding
and saturation effects.}
140 infrared excess sources, among them
23 Class~I sources. 
Another large fraction of the known YSO population is located in 
an elongated cluster
at the southern edge of S256, and there are several
smaller clusters at different 
locations (see Fig.~12 in \citealt{cha08}).

About 
1$'$ north of the center of S255-IR, a strong far-infrared source, S255-N,
 was detected by \citet{jaf84}. This object is also associated
with massive star formation and it is believed to be at an even 
earlier evolutionary stage than S255-IR  
\citep{kur94,kur04,cyg07,wan11}. Another deeply embedded region, S255-S \citep{wan11},
is located about 1$'$ south-west from 
 S255-IR.
It exhibits strong mm continuum emission \citep{wan11,dif08} but no other sign
of active star formation in near- and mid-IR observations. 
\citet{min07} suggested that this sub-region is in a
very early pre-stellar phase of evolution. 
Scenarios for the spatial and temporal sequence of star formation in the
S254-S258 complex have been recently discussed in \citet{cha08},
\citet{bie09}, and  \citet{wan11}.

The distance of the S254-S258 complex was only poorly known until recently.
\citet{pis76} and \citet{mof79} derived a value of 2.5~kpc; similar values
\citep[e.g.,][]{cha08}, but also lower values down to 1.5~kpc were used in later studies.
\citet{ryg10} performed high-precision astrometry using the
6.7~GHz methanol maser emission from the source J0613+1708 in S255-IR
and derived a very accurate trigonometric parallax of $(1.59 \pm 0.07)$~kpc
for S255.  We will use this new and reliable distance for our present study.

\subsection{Evolution scenarios for the central region S255/257}

The most remarkable part of the S254-S258 complex and the
focus of the study presented here is the central
region around the two \ion{H}{II} regions S255 and S257.
A very interesting, as yet unexplained, feature of this region is
that the two B0 stars exciting these \ion{H}{II}
regions, ALS~19 and HD~253327, appear to be more or less isolated and do not 
have any obvious co-spatial low-mass clusters \citep{zin93}. This is remarkable because, 
according to the standard field star IMF  \citep[e.g.,][]{kro01} each B0 star 
($M_\ast \approx 15\,M_\odot$) should be accompanied by about 300
lower-mass stars. Fundamentally different possible explanations for the 
apparent absence of low-mass stars around these B0 stars
have been proposed over the years.

One scenario is based on the assumption of {\em bimodal star formation}. It assumes
that the two high-mass B0 stars formed independently from the low-mass young stars
in the central cluster and in a fundamentally different processes. The lack of clusters
of low-mass stars around these two B0 stars would then 
imply that these high-mass stars formed in isolation \citep[see][]{zin93}. 
One problem with this scenario is that in
basically all other well investigated star forming regions, high-mass stars 
are always associated with large numbers of low-mass stars 
\citep{testi99,bri07}. The case of S255 and S257 would then represent a quite 
unique exception if the absence of low-mass stars was confirmed.

The second scenario assumes that the two B0 stars formed together with the 
low-mass
stars in the dense central cluster S255-IR, but were {\em dynamically ejected}, e.g.~by means of
close stellar encounters  and N-body interactions.
This model predicts that both B0 stars should move away from the central clusters
with substantial velocities. Unfortunately, 
the available
HIPPARCOS proper-motions for the stars are not accurate enough to either support or rule out this scenario.

 A third scenario assumes 
{\em multiple stellar generations and triggered star formation}.
Here, the two B0 stars belong to an earlier generation of stars that
formed several Myr ago in this area.
The expanding \ion{H}{II} regions swept up diffuse gas and dust in 
their surroundings into shells, and
formed the dense cloud in the interaction zone between them. 
This process of creating new clouds at the edges of shells or bubbles driven by high-mass stars
is well established and observed at many locations \citep[see, e.g.,][]{Brand11,Zavagno10,Deharveng09}.
The particularly strong compression of the cloud 
at the intersection of the two shells, caused 
by the ongoing expansion of the \ion{H}{II}
regions, may have triggered the formation of a new generation of stars, i.e.~the embedded 
cluster of young stellar objects. The ongoing expansion of the interacting
bubbles would also provide a natural explanation why the youngest regions
S255-N and S255-S are found just above and below the central young cluster
S255-IR at the intersection of the shells.

\subsection{Importance of the low-mass stellar population}

A discriminant between the different evolution scenarios for the
S255/257 region is the presence or absence of 
low-mass stars associated with the two B0 stars. While no low-mass
stars should be present in the case of the bimodal star formation model 
or the dynamical ejection model, the
multi-generation model predicts the presence of several hundred low-mass stars near these B0 stars, 
since the stellar
populations in basically all well-investigated OB associations follow the standard field star
IMF \citep{bri07}. A
handful emission line stars and a few dozen infrared excess objects 
\citep[see][]{cha08} are known inside or near the two \ion{H}{II} regions, 
but their
numbers are far to small for the expected low-mass population associated with the massive B0 stars. 

The apparent lack of associated low-mass stars may, however, just be a result
of the sensitivity limits of the existing observations:
a population of several Myr old low-mass stars would be quite hard to 
identify in the present optical
and infrared images, for several reasons. First, the low-mass stars would not be densely clustered around the B0
stars but scattered over a rather wide area, up to $\sim 10$~pc away from the massive stars, as typical for
subgroups in OB associations. Second, these low-mass stars would be quite hard to see in most existing optical or
infrared images of the region: since a $\sim$ 10 Myr old 1$M_{\odot}$ [0.2$M_{\odot}$] stars should have magnitudes
of V $\geq$ 18.3 [22.2] and K $\geq$ 14.7 [17.0], they would be not easy to detect in the nebulosity of the
\ion{H}{II} regions and the diffuse infrared emission in this region. Third, even the availability of Spitzer
observations is of limited use here: although Spitzer data are sensitive 
enough to detect a good fraction of the low-mass stars,
the Spitzer images of the region are dominated by unrelated field stars (note that this region lies very close to
the galactic plane). The usual approach to identify young stars by their infrared excesses is not feasible here,
because at an age of more than a few Myr, most of the low-mass association members have already lost their
circumstellar disks and thus should not exhibit infrared excesses \citep{bri07}.
\textit{It is thus impossible to identify
and distinguish a population of several ($\sim 3 - 10$)~Myr old low-mass association members from
unrelated field stars with optical or infrared photometry alone.}

Sensitive X-ray observations can provide
a very good solution of this problem, since they allow to
detect the young stars by their strong X-ray emission
\citep[e.g.,][]{fei07}
and efficiently
discriminate them from the numerous
older field stars in the survey area.
The median X-ray luminosity of $\la 10$~Myr old solar-mass stars is 
$\approx 10^{30.4}\,\rm erg\,s^{-1}$; this is nearly 1000 times higher than for solar-mass field stars \citep[see][]{pre05}, and makes these
young stars relatively easily detectable X-ray sources. Another very important
aspect is that X-ray observations trace magnetic activity rather than photospheric or circumstellar disk emission
from young stars, and are thus complementary to the available optical and infrared data of the region. The X-ray
selected sample of low-mass stars will be not biased toward stars with circumstellar disks identified in the
Spitzer data. Furthermore, an X-ray image is not subject to confusion from bright diffuse emission 
by heated gas
and dust. X-rays can penetrate deeply into obscuring material and are very effective in detecting 
embedded YSOs \citep{get05b}.
Many X-ray studies of star forming regions have demonstrated
the success of this method
\citep[see, e.g.,][]{PZ02,bro07,FP07,tow11}.
Also, the
 relations between the X-ray properties and basic stellar
properties in young stellar populations
are now  very well established from very deep X-ray
observations such as the
{\it Chandra} Orion Ultradeep Project (COUP)
\citep[see][]{get05,pre05b}.
To summarize, a deep X-ray image of the S254-S258 complex can reveal the full young stellar populations
in the area and provide essential information about the star formation history.

At distances beyond 1~kpc, very good angular resolution is required
to resolve the individual sources in the dense young clusters 
and to allow a reliable identification of the X-ray sources with the
numerous infrared sources (note that the complex is almost exactly on the galactic plane, $b = -0.048\degr$).
The {\em Chandra} X-ray observatory, that provides an on-axis PSF of $\leq 1''$,
 is the only currently active X-ray mission that has
sufficient angular resolution for this purpose.

We have therefore performed a deep {\em Chandra} X-ray observation of this extraordinary
star forming region in order to uncover the population of low-mass association members. 
Our study focuses on the central region of the S254-S258 complex,
i.e.~the two \ion{H}{II} regions S255 and S257 and the
embedded cluster S255-IR between them.
A characterization of the size, the spatial distribution, and the
properties of the low-mass population can provide important information
on the star formation history and discern between the different models for
the relation between S255/S257 and the embedded cluster S255-IR.
In Section \ref{sec:obsdat} we describe the \textit{Chandra} observations and data reduction. 
Section 3 presents the basic X-ray properties of the
detected sources. Section \ref{sec:pop} analyzes the X-ray population of the 
S254-S258 star forming complex,
and Section 5 discusses the spatial distribution of the X-ray sources and
the implications on the star formation process in S254-S258.
A more detailed analysis of the optical and infrared
properties of the individual X-ray detected young stars
(that can provide direct information on the
ages, masses, and the circumstellar disks around these
stars) will be presented in a forthcoming paper.


\section{Observations and data reduction}\label{sec:obsdat}

   \begin{figure*} 
\sidecaption 
 \parbox{12cm}{\includegraphics[width=11.9cm]{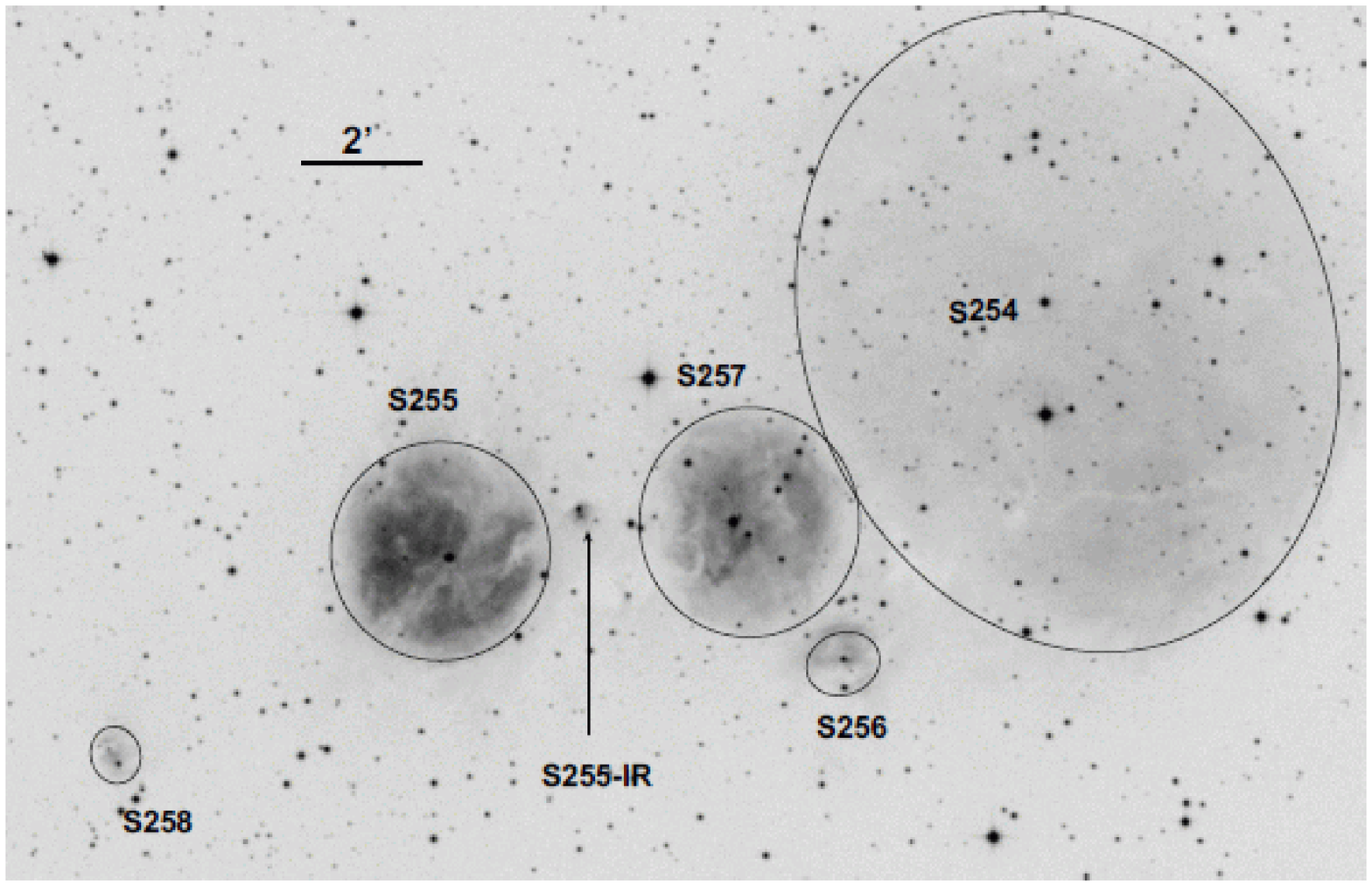}\\
    \includegraphics[width=11.9cm]{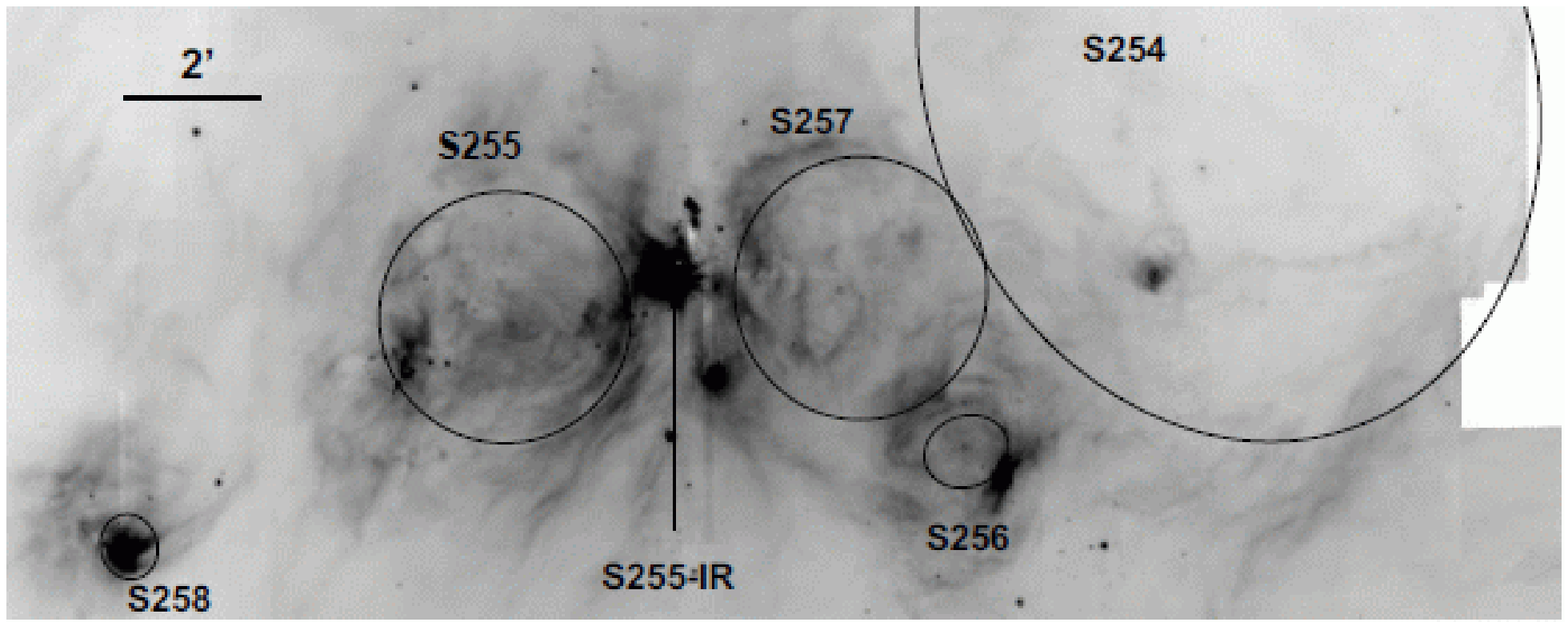}\\
	\includegraphics[width=12cm]{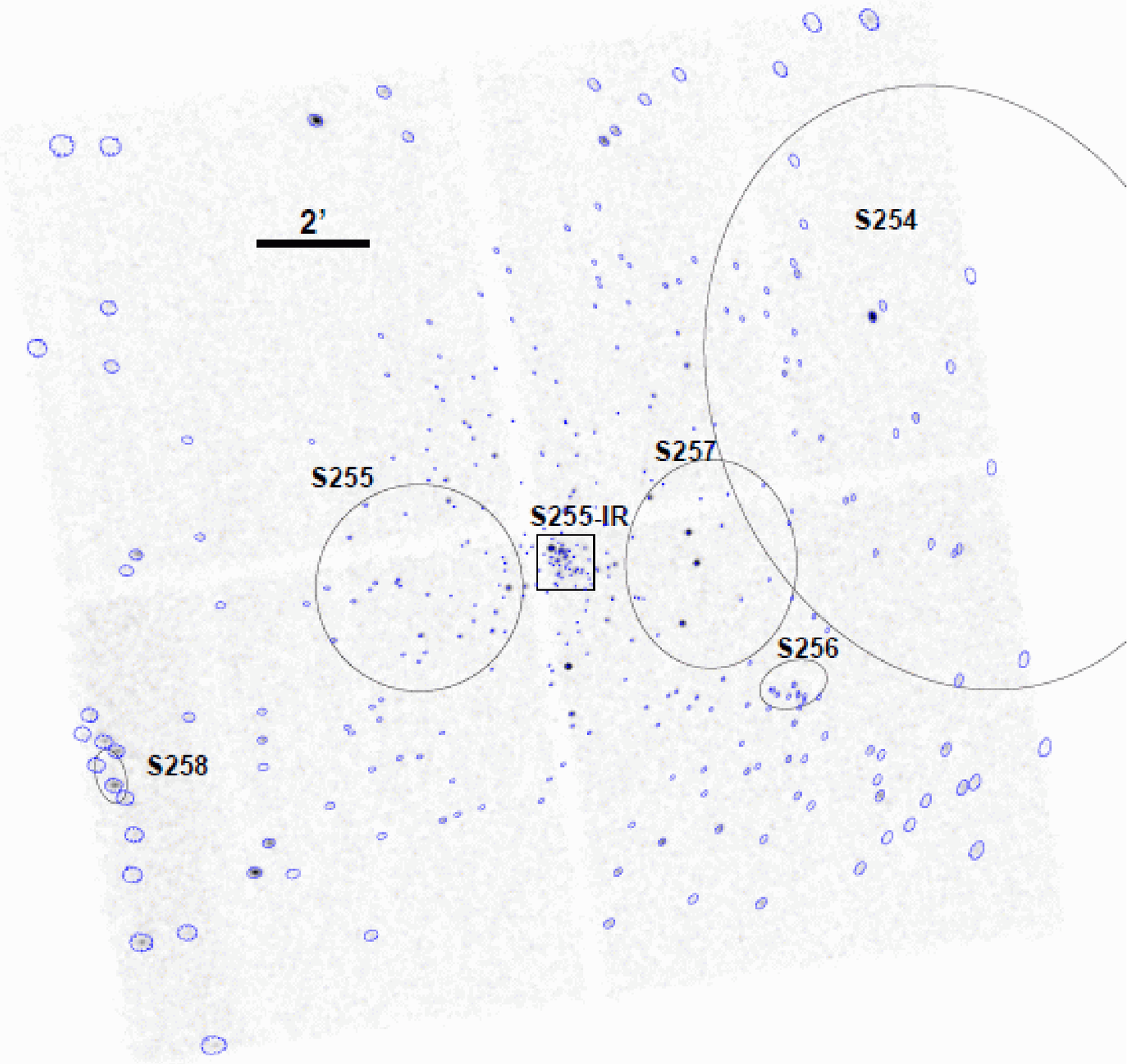}}
   \caption{\textbf{Top:} Negative grayscale representation
of the optical image of the S254-S258 complex from the Digitized
Sky Survey. The black ellipses
   represent the five \ion{H}{II} regions that define the complex.
The location of the central, optically invisible embedded cluster 
S255-IR is marked by the arrow.\newline
\textbf{Center:} \textit{Spitzer}~IRAC~4 image of the central part
of the S254-S258 complex. This image was created from the
 basic calibrated data products for the
programs 201 and 30784 retrieved from the
\textit{Spitzer} archive and  mosaicked with the MOPEX
software available from the Spitzer Science Center.
Note that parts of the bright emission from the central embedded cluster 
S255-IR is saturated in these data.\newline
\textbf{Bottom:} \textit{Chandra} ACIS-I image of S254-S258 in the 
[0.5--8.0]~keV band. Blue ellipsoids represent extraction regions
   for the individual detected X-ray sources based on a model of the local PSF that
encircles 90\% of total energy.}
              \label{fig:allsrc}%
    \end{figure*}

\subsection{\textit{Chandra} X-ray observations of S254-S258}

The S254-S258 complex was observed (PI: Th.~Preibisch)
in November 2009 with the 
Imaging Array of the \textit{Chandra} 
Advanced CCD Imaging Spectrometer
(ACIS-I).
ACIS-I provides a field of view of $17' \times 17'$ on the sky.
At the 1.6~kpc distance of S254-S258 this corresponds to $7.9 \times 7.9$~pc.
The aimpoint of the observation was
  $\alpha ({\rm J2000}) = 06^{\rm h}\,12^{\rm m}\,54.0^{\rm s}$,
$\delta = +17\degr\,59'\,24''$. The observation was performed in the standard
``Timed Event, Faint'' mode (with $3 \times 3$ pixel event islands).
The total net exposure time of 74\,725~s (20.76 hours) was split 
into two parts, separated by about 4 days.
The details of these two observation parts are given in Table \ref{tab:obslog}. 
Since the roll angles (i.e.~the orientation of the detector on the sky) is equal for both observations,
each source is at the same detector position in both parts, making
merging of these two data sets rather straightforward.
Two of the CCDs of
the ACIS-S spectroscopic array were also turned on during our observations.
However, since the PSF at the large off-axis angles at these detectors
is strongly degraded, their point-source sensitivity is reduced; only
six X-ray sources are detected in the field of these ACIS-S chips.

The basic data products of our observation are the two 
Level 2 processed event list
provided by the pipeline processing at the  \textit{Chandra} X-ray Center,
that list the arrival time, location on the detector and energy for 
each of the 626\,140 detected X-ray photons. 
We combined the two pointings with 
the {\it merge\_all} script, a {\em Chandra} contributed software that 
make use of standard {\it CIAO}\footnote{Chandra Interactive Analysis of Observations, version 4.2:\\ 
http://cxc.harvard.edu/ciao/index.html} tools. 
The mean background count rate in our merged image, determined from 
several large source-free regions, is 
$2.95 \times 10^{-7}$~counts~s$^{-1}$~pixel$^{-1}$, corresponding
 to a mean background level
of 0.02~counts~pixel$^{-1}$. %

\begin{table*}
\centering                          
\caption{{\em Chandra} observation log}             
\label{tab:obslog}      
\begin{tabular}{c c c c c}        
\hline\hline                 
Obs.Id. & Date & Start -- End Time [UT]& Exposure time &  Level 2 events \\    
\hline                        
   10983 & 2009-11-16 &11:40:48 -- 23:35:08& 40\,570 s&  340\,411\\	 
   12022 & 2009-11-20 &05:14:15 -- 15:18:20& 34\,155 s&  285\,729 \\	 
\hline                                   
\end{tabular}
\end{table*}

At a distance of 1.6 kpc, the expected ACIS point source sensitivity limit
for a 5-count detection on-axis in a 75~ks observation is
$L_{\rm X,min} \sim  10^{29.5}$~erg~s$^{-1}$,
assuming an extinction of 
$A_V \leq 2.5$~mag ($N_{\rm H} \leq 5 \times 10^{21}\,{\rm cm}^{-2}$)
as typical for the stars in the \ion{H}{II} regions,
and a thermal plasma
with $kT = 1$~keV \citep[which is a typical value for young stars; see, e.g.,][]{pre05b}.
Using the  empirical relation between X-ray luminosity and stellar mass
and the temporal evolution of X-ray luminosity
from the 
sample of young stars in the Orion Nebula Cluster
that was very well studied in the \textit{Chandra} Orion 
Ultradeep Project \citep{pre05b,pre05},
we can expect to detect  almost
all stars in S254-S258 with masses greater than $0.5\,M_\odot$
and about half of the $0.1-0.5\,M_\odot$ stars.
The expected level of detection completeness is $\ga 90\%$ 
for stars with $M_\ast \ge 0.5\,M_\odot$ (corresponding approximately to 
spectral types earlier than M1) and drops 
below 50\% at $M_\ast \le 0.25\,M_\odot$ (spectral types $\la$~M5).
Note that these values are valid for the central part of the
observed field; 
sensitivity is $\sim 3-4$ times worse 
at the edges of the ACIS field.

\subsection{Source detection and X-ray source catalog}

The source detection was performed in a two-step process. The
first detection step was performed in a rather aggressive manner
in order to find even the weakest possible sources, deliberately accepting 
some degree of false detections. In the second step, this list of 
potential sources was then cleaned from spurious detections
by a detailed individual analysis.
We employed the {\sc wavdetect} algorithm
\citep[a {\it CIAO}
mexican-hat wavelet source detection tool]{fre02}
for  locating X-ray sources in our merged image,
and used a rather low detection threshold of 10$^{-5}$.
This step was performed in three different energy bands, 
the total band [0.5-8.0]~keV, the soft band [0.5-2.0]~keV, and the 
hard band [2.0-8.0]~keV, and with wavelet scales between 1 and 16 pixels. 
We also performed a visual inspection of the images and added some 
30 additional candidates to the merged catalog from the wavelet analysis,
resulting in a final catalog of 511 potential X-ray sources.

To clean this catalog from spurious sources, we then performed a detailed
analysis of each individual candidate source with the
ACIS Extract (AE hereafter) software
package\footnote{http://www.astro.psu.edu/xray/docs/TARA/ae\_users\_guide.html} 
\citep{bro10}. A full description of the 
procedures used in AE can be found in \citet{get05}, \citet{ton03} 
and \citet{bro07}. 
The following three steps were performed by AE in order to prune our input catalog from spurious detection
(including afterglows):
\begin{enumerate}
\item{Extraction regions were defined as the 90\% contours of the local PSF 
(or smaller in the case of other nearby sources), and
source events were extracted. Energy dependent corrections for the finite 
extraction regions were applied;}
\item{Local background events were extracted after masking all the sources in the catalog;}
\item{The Poisson probability ($P_{B}$) associated with the ``null hypothesis'',
i.e.~that no source exist 
and the extracted events are solely due to
Poisson fluctuations in the local background, is computed for each source.}
\end{enumerate}

All candidate sources with $P_{B} >0.01$ were rejected as background fluctuations.
After 8 iterations of this pruning procedure our final
catalog consisted of 364 sources.
It contains 344 primary sources with 
$P_{B} < 0.003$, and 20 tentative sources with $0.003 < P_{B} < 0.01$.
The extraction regions for the sources in our final catalog
are plotted on the {\em Chandra} image in Fig.~\ref{fig:allsrc}.

\subsection{X-ray point-source analysis with \textit{ACIS Extract}}

The AE software also determines basic properties for each of the
detected sources, such as the net (i.e., background-subtracted) counts
in various energy bands, the median photon energy,
 statistical test for variability, and a measure of the incident 
photon flux.
These properties are reported in Table~2 
\onltab{2}{}
(available in the electronic edition). Sources are sorted
by increasing right ascension and identified by their sequence number (Col.~1)
or their IAU designation (Col.~2).
While the general X-ray properties were determined from the merged data set
(the AE software is well suited for this purpose),
we note that the spectra (see Sect.~\ref{ssec:spec}) 
were extracted from the individual observations.

\subsection{Expected contamination of the X-ray source sample}
\label{ssec:contamination}

As in any X-ray observation,
there must be some degree of contamination
by galactic field stars as well as extragalactic sources.
To quantify the expected level of this contamination,
we consider the results from the recent
\textit{Chandra} Carina Complex Project \citep[CCCP; see][]{tow11},
for which the
individual pointings had very similar exposure times ($\approx 60 - 80$~ks)
as our S254-S258 pointing.
Furthermore, S254-S258 is at nearly the same galactic latitude as the Carina Nebula,
suggesting that the background contamination should be very similar in
these two regions.

For the CCCP data set, the
classification study of
\cite{bro11b}, which considered the
X-ray, optical, and infrared properties of the sources
(that differ for the different contaminant classes),
found that
716 X-ray sources in the 1.46 square-degree CCCP survey are
are foreground stars, 16 are background stars,
and 877 are extragalactic (AGN) contaminants.
Scaling these numbers to the field-of-view of our S255 pointing
gives 39 foreground stars, 1 background star, and 48
extragalactic (AGN) contaminants.
However, since S254-S258 is considerably closer (1.6~kpc) than
the Carina Nebula (2.3~kpc), the number of foreground stars should be
accordingly smaller, approximately by a factor of $(1.6/2.3)^3 \approx 0.34$.
Furthermore, the number of foreground stars
in the Carina Nebula is particularly high
since this direction is close to the tangent point of the
Carina spiral arm.
These considerations imply that the contamination in our S254-S258 field
should be
clearly dominated by $\sim 48$ expected extragalactic sources (AGNs).

A characteristic of extragalactic contaminants is that their
optical and infrared counterparts should be very faint, in
fact mostly undetected in the available optical and infrared images.
As we describe in more detail below, our search for
counterparts of the X-ray sources left
46 X-ray sources outside the central embedded cluster S255-IR without
optical or infrared counterparts.
This number agrees well with
the expected number of extragalactic contaminants.

Assuming at most 10 contaminating foreground/background stars, the
total expected number of contaminants would be $\la 58$.
With a total number of 364 detected X-ray sources, the expected
level of contamination for our sample is thus $\la 15\%$.

\section{Properties of the X-ray source in S254-S258}

\subsection{X-ray fluxes and luminosities}\label{ssec:lum}

An accurate determination of the intrinsic X-ray source luminosities
requires good knowledge of the X-ray spectrum. However,
for the majority of the X-ray sources the number of detected
photons is too low for a detailed spectral analysis.
Only 25 sources in our catalog have more than 80 net counts, the
practical lower limit for meaningful spectral analysis.
For these bright sources we performed a detailed
spectral fitting analysis to derive the plasma temperature and
the extinction, and from these quantities we can calculate the
intrinsic (i.e.~, extinction-corrected) X-ray luminosities,
as described in detail in Section \ref{ssec:spec}.

For the weaker X-ray sources, for which a meaningful spectral
analysis is not feasible, one cannot determine intrinsic
X-ray luminosities without knowledge of the extinction.
This is a substantial problem because the
young stars in the S254-S258 complex show a very wide range
of extinctions. There are numerous optically visible stars with low
obscuration (at most a few magnitudes of visual extinction), while
other stars suffer from cloud extinction up to about 
$A_V \sim 20$~mag, and embedded YSOs show additional 
circumstellar extinctions up to
$A_V \sim 50$~mag and beyond \citep{cha08}.
This implies that we cannot simply use a common count-rate to flux
conversion factor to determine intrinsic X-ray luminosities but have
to consider each source individually.

\subsubsection{Observed X-ray fluxes}
\label{ssec:flux2}

An estimate of the \textit{observed} (i.e.~\textit{not} the intrinsic)
X-ray flux is computed by AE. This quantity, called $FLUX2$,
is calculated from the number of detected photons
and using a mean value of the instrumental effective area (through the
Ancillary Response Function, ARF) over energy.
The $FLUX2$ values (derived for the full band, i.e.~$[0.5-8]$~keV range),
are reported in
column (3) in Table~3 (available in the electronic edition).
\onltab{3}{}
It should be noted that
this $FLUX2$ values suffer from a systematic error with respect to the true
incident flux, because the use of a mean ARF is only correct in the
hypothetical case of a flat incident spectrum, an assumption that
probably not fulfilled.
Nevertheless, the $FLUX2$ represents the best flux estimate
that can be obtained  for weak sources.

   \begin{figure}
   \centering
\includegraphics[width=8.8cm]{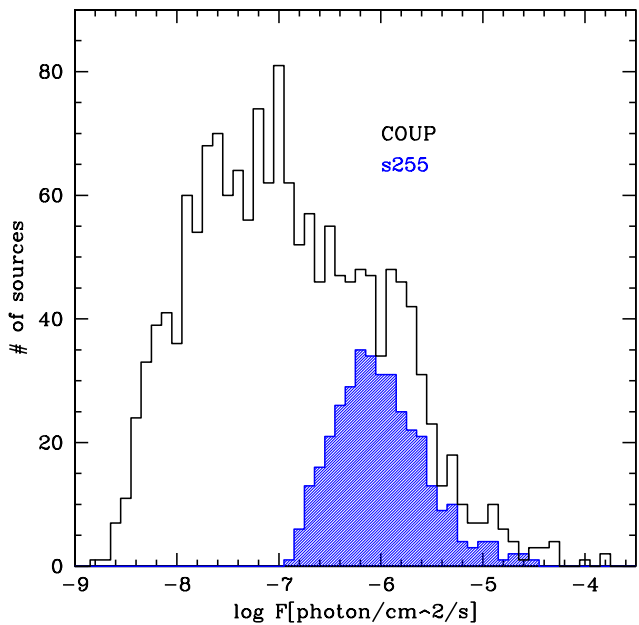}
   \caption{The blue histogram shows the distribution of FLUX2 values for the X-ray sources
in S254-S258. For comparison, the distribution of fluxes from the COUP data from \citet{get05} is 
shown by the black histogram,
	      scaled to the distance of S254-S258, i.e.~1.6~kpc.} \label{fig:flf}%
    \end{figure}
In Figure \ref{fig:flf} we compare the distribution of $FLUX2$ values for
the S254-S258 sample to that of the stars in the Orion Nebula Cluster
obtained in the context of the \textit{Chandra} Orion Ultradeep Project
\citep[COUP, see][]{get05}. Note that the COUP fluxes were scaled to the
1.6~kpc distance of S254-S258.
The two distribution show a very similar shape in the range
between $\approx 10^{-6.0}$~photons~cm$^{-2}$~s$^{-1}$ and
 $\approx 10^{-4.5}$~photons~cm$^{-2}$~s$^{-1}$.
The differences between the two distributions can be explained as follows:

First, the COUP sample shows a few stars with 
fluxes of $> 10^{-4.5}$~photons~cm$^{-2}$~s$^{-1}$, while no such
very bright sources are
seen in S254-S258. These very bright sources
are the highly X-ray luminous O-type stars in the Orion Nebula Cluster. 
The fact that the S254-S258 does not
contain such high-mass stars explains the absence of similarly high values
in the observed distribution of incident fluxes for S254-S258.

Second, the peak and turn-over of the S254-S258 distribution at
 $\approx 10^{-6.2}$~photons~cm$^{-2}$~s$^{-1}$
is a direct consequence of the higher sensitivity limit
of our S254-S258 X-ray observation. As S254-S258 is about 4 times
more distant than the ONC, and since the exposure time of
our S254-S258 \textit{Chandra} observation is less than one tenth of the
840~ks COUP observation,
the expected sensitivity limit should be about 150 times higher.

Third, the number of sources per bin is always lower for 
S254-S258 compared to the ONC. This suggests that the total stellar
population in the observed part of S254-S258 is smaller than in the
ONC as observed in the COUP.

\subsubsection{X-ray luminosities from \textit{XPHOT}}
\label{ssec:xphot}

An estimate of the intrinsic, i.e.~extinction
corrected, X-ray luminosity for sources that are too weak for a
detailed spectral analysis can be obtained with the
\textit{XPHOT} software\footnote{http://www.astro.psu.edu/users/gkosta/XPHOT/},
developed by \citet{get10}.
\textit{XPHOT} is based on a non-parametric method for the calculation 
of fluxes and absorbing X-ray column densities of weak
X-ray sources. X-ray extinction and intrinsic flux are estimated from
the comparison of the apparent median energy of the
source photons and apparent source flux with those of high signal-to-noise
spectra that were simulated using
spectral models characteristic of much brighter sources of similar class 
previously studied in detail. This method requires at least 4 net counts
per source (in order to determine a meaningful value for the
median energy) and can thus be applied to 255 of our 364 sources.
Columns (4) to (7) of Table~2 
report  apparent and intrinsic (corrected for absorption, noted with subscript ${c}$) luminosities in
the hard and total band, assuming a distance of 1.6~kpc.
The resulting intrinsic X-ray luminosities range from $10^{29.4}$ to
$10^{32.3}$~erg~s$^{-1}$.

   \begin{figure*}
   \centering
   \includegraphics[width=8.0cm]{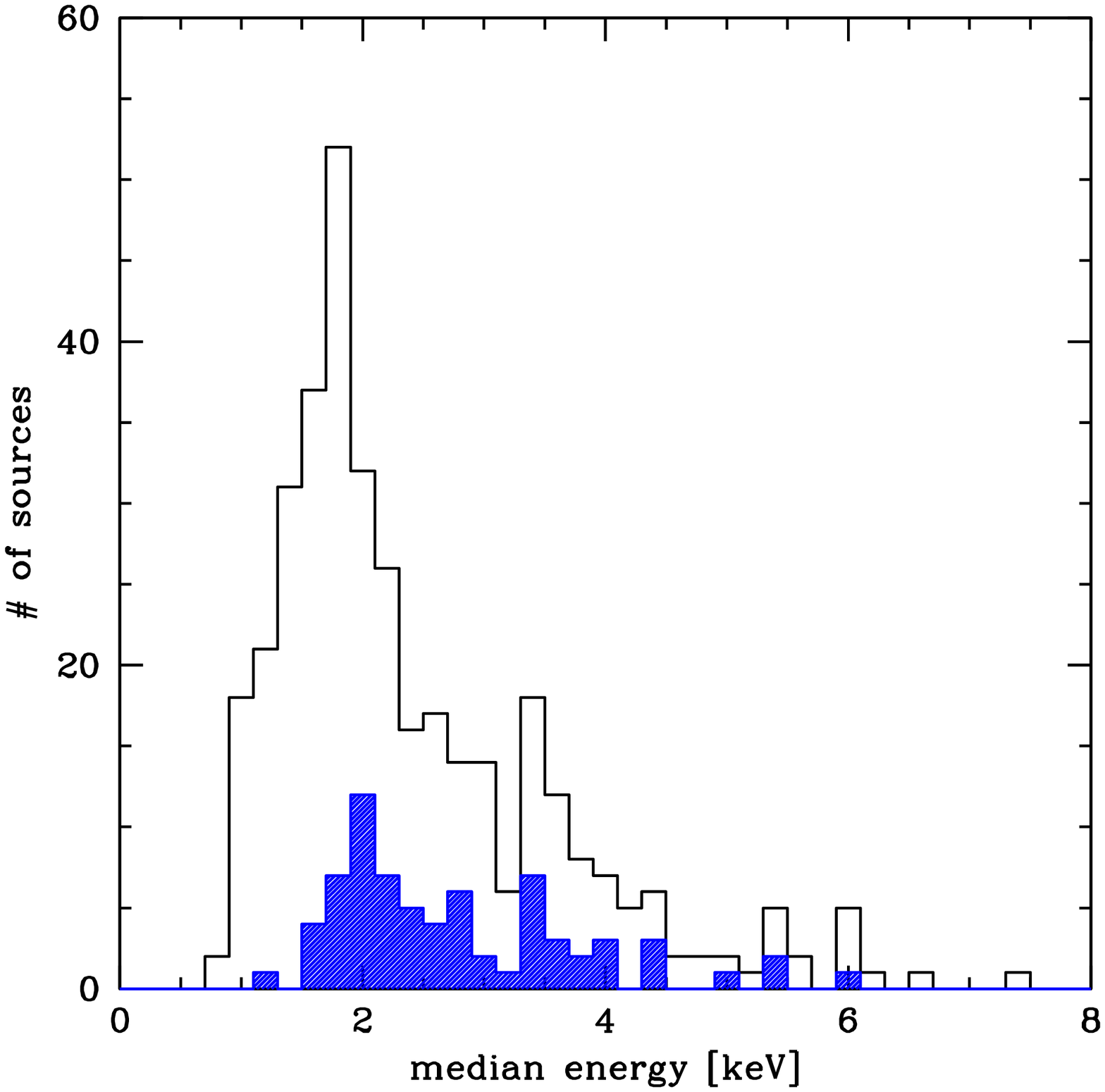}
   \includegraphics[width=8.0cm]{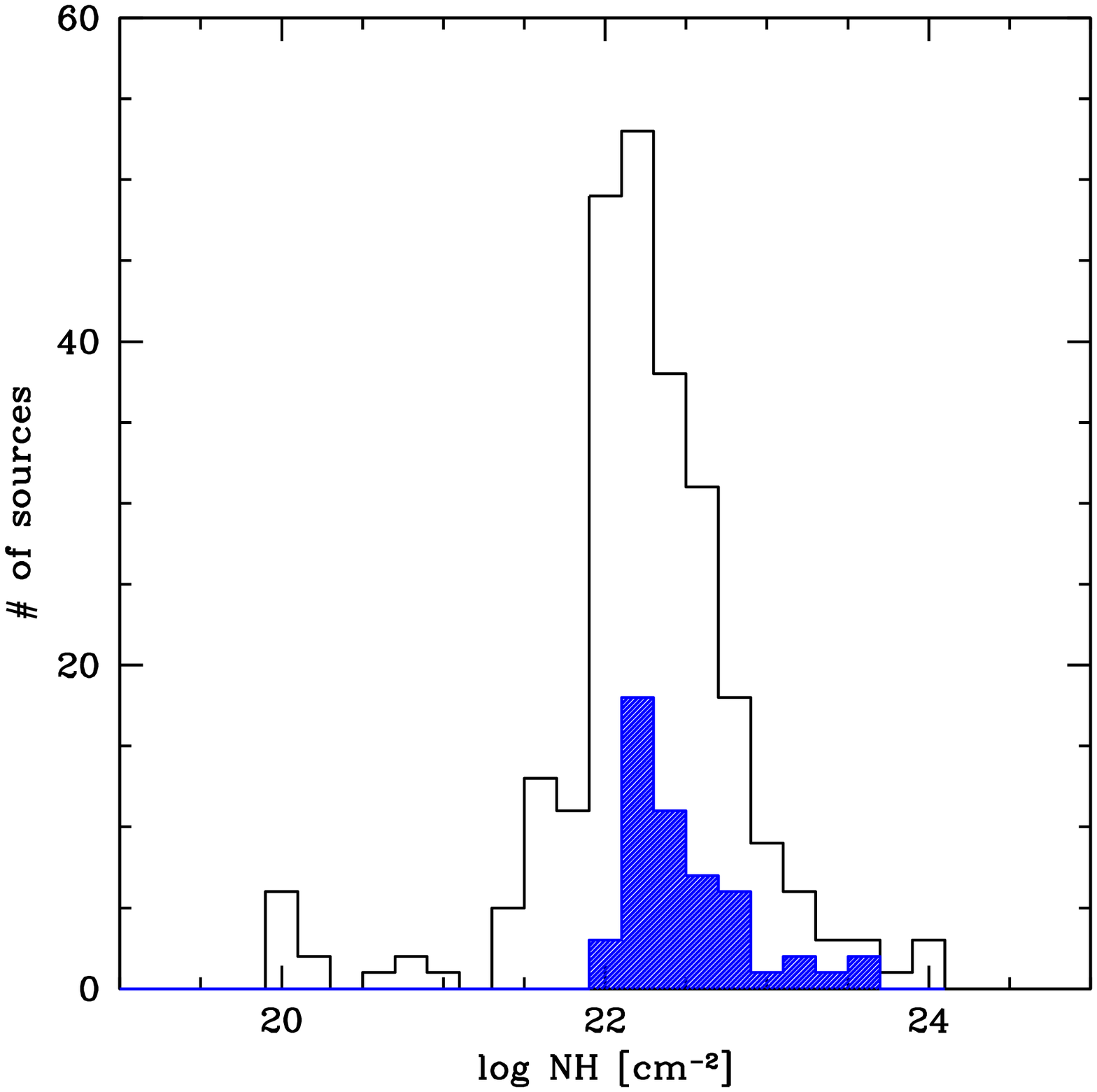}
   \caption{Distributions of median photon energies (left) and 
hydrogen column densities (right) determined by {\it XPHOT} (right) for the whole 
sample (black histograms).  The blue dashed histograms show the
distributions restricted to the sub-sample of sources
located in the 1 arcmin
   radius region centered on the embedded cluster S255-IR.}
              \label{fig:dist2}%
    \end{figure*}

Figure~\ref{fig:dist2} shows the distribution of median photon energies
and the deduced hydrogen column densities estimated by {\it XPHOT}.
The median value of the derived hydrogen colum densities
is $\log \left( N_H\,[{\rm cm^{-2}}] \right) =  22.04$,
 corresponding to a visual absorption of $A_{V} \sim 6$ magnitudes.
If we consider the sub-sample of sources located in the
central embedded cluster S255-IR, this value rises to
$\log \left( N_H\,[{\rm cm^{-2}}] \right) =  22.21$
($A_{V} \sim 9$ magnitudes), clearly showing stronger
obscuration for the embedded sources.

\subsubsection{X-ray spectral fits of bright sources} 
\label{ssec:spec}

   \begin{figure*}
   \centering
\includegraphics[angle=-90,width=8.5cm]{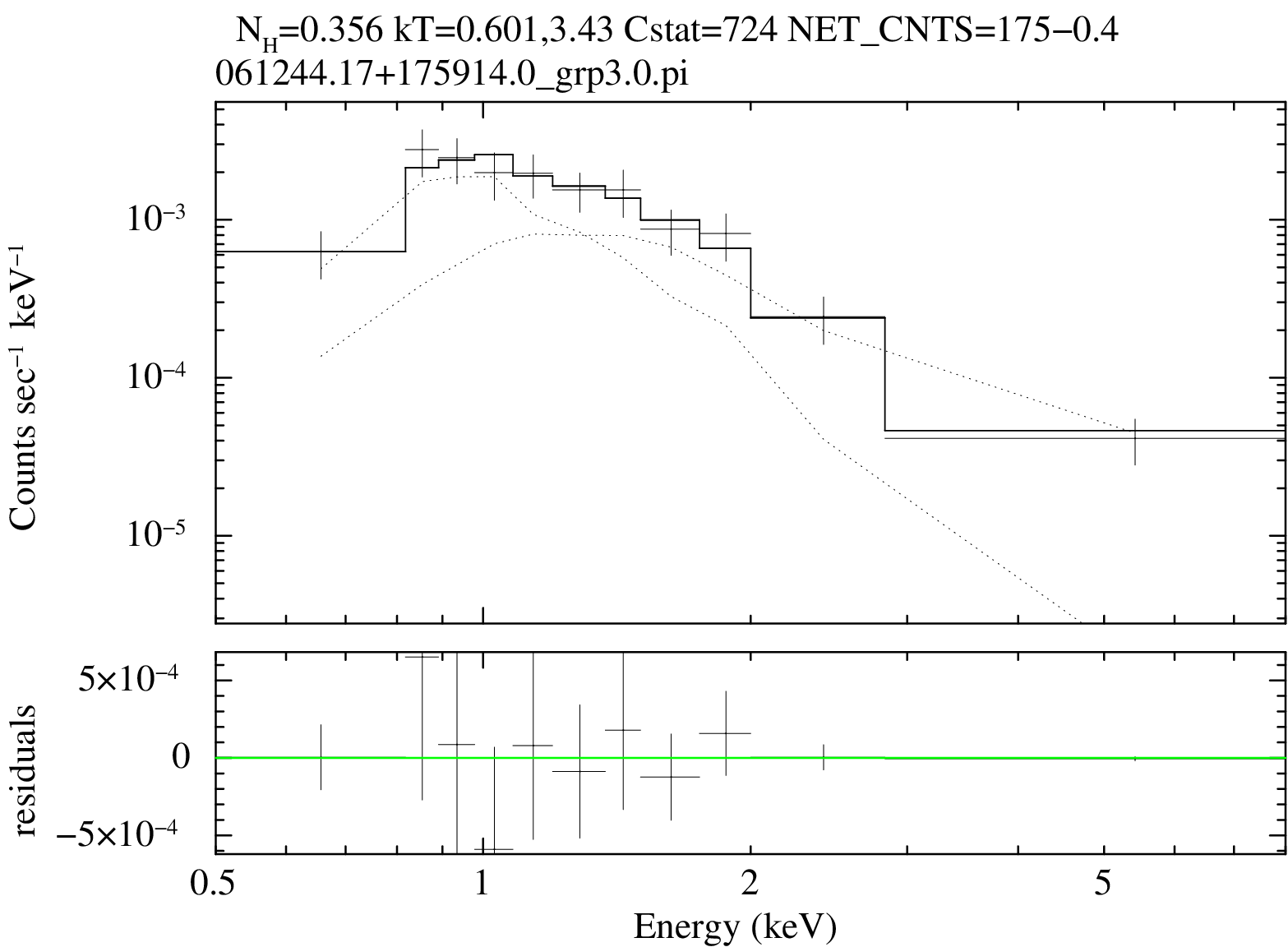}
\includegraphics[angle=-90,width=8.5cm]{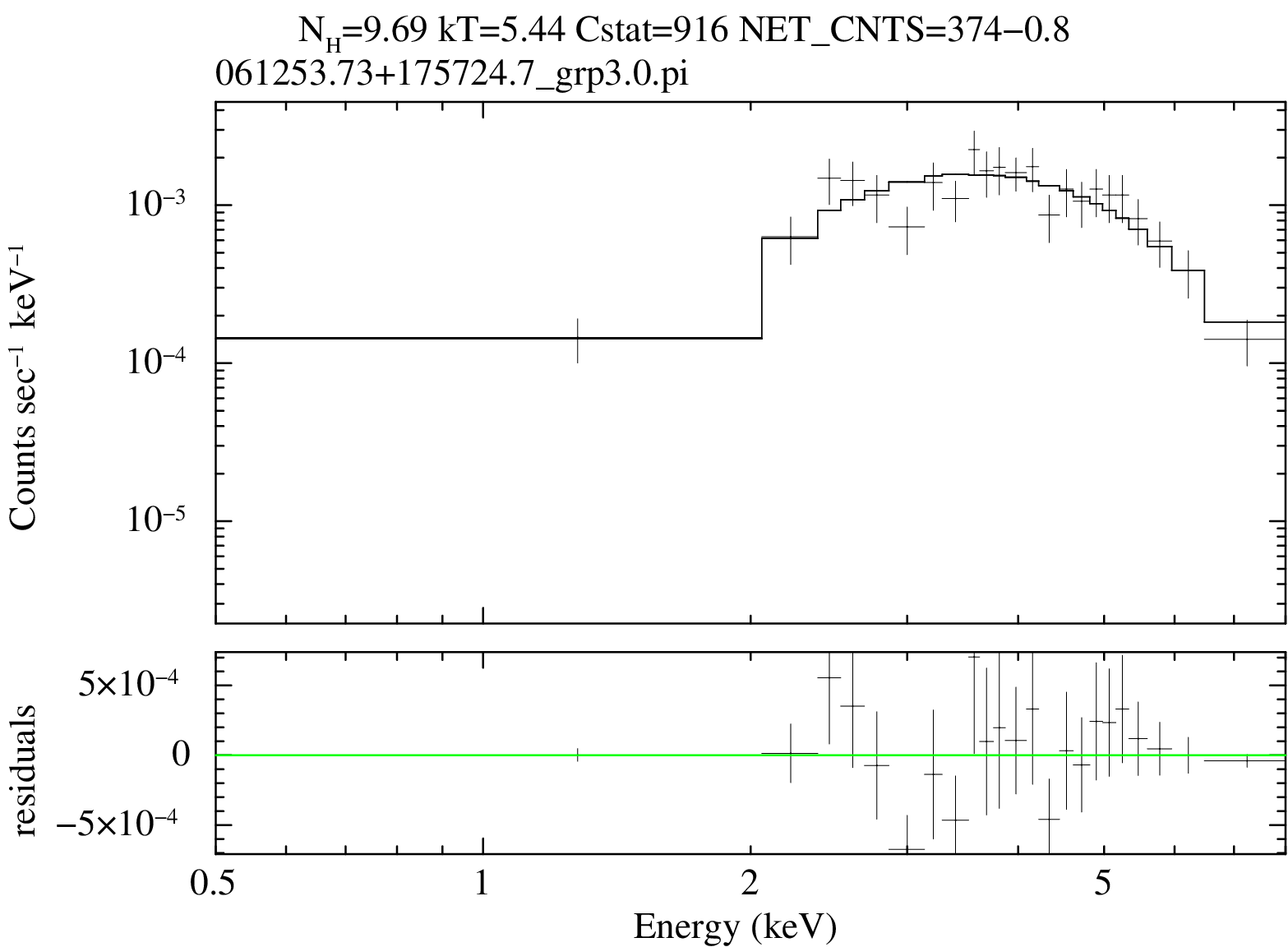}\\
\includegraphics[angle=-90,width=8.5cm]{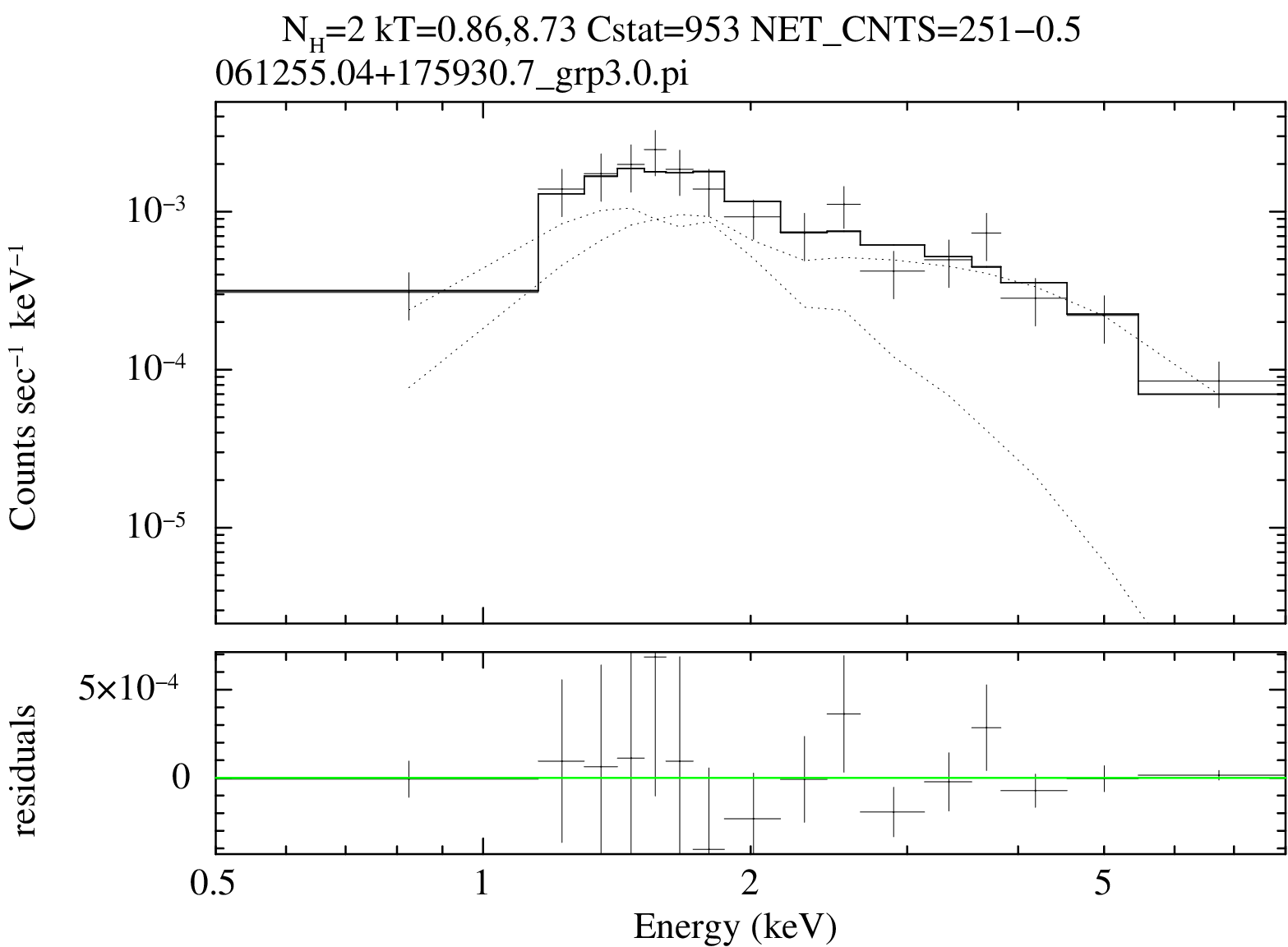}
\includegraphics[angle=-90,width=8.5cm]{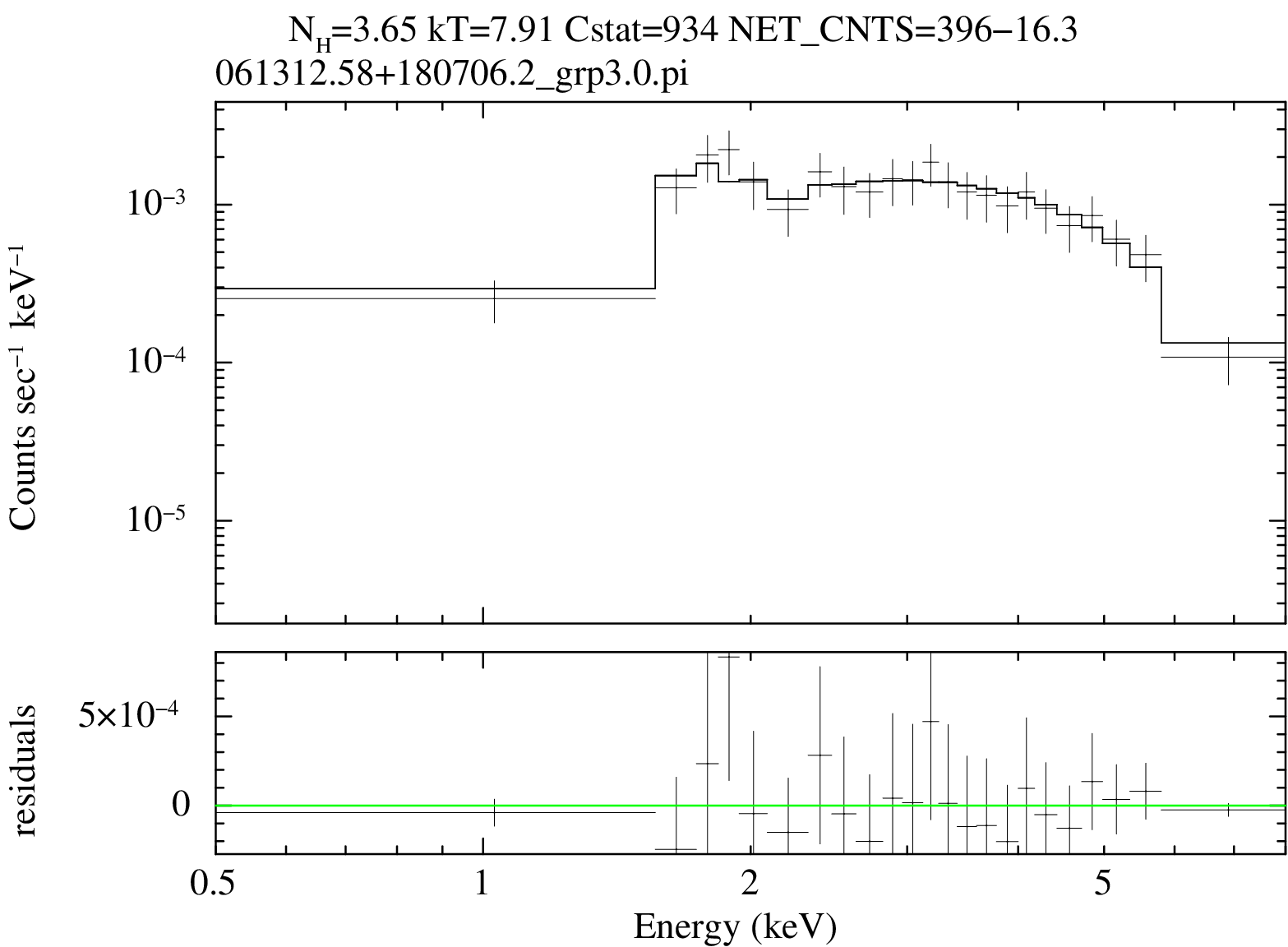}
   \caption{\textit{Chandra} X-ray spectra and best-fit models
of four bright X-ray sources in S254-258.
The crosses show the measured spectra, the solid lines show the best-fit
models; in the two cases where a two temperature model was required, the 
dotted lines show the two individual spectral model components. 
CXOU~J061244.17+175914.0 is the B0.5 star HD 253327 that illuminates the S257 nebula.
CXOU~J061253.73+175724.7 is the source with the highest derived X-ray luminosity
in our sample; it is located in the embedded cluster S255-IR. 
CXOU~061255.04+175930.7 is another embedded infrared source in S255-IR.
CXOU~061312.58+180706.2 is the source with the second highest X-ray luminosity
in our sample.}
              \label{fig:spec}%
    \end{figure*}
For the 
25 sources in our sample with more than 80 net counts we performed a 
spectral fitting analysis
using AE and the {\it XSPEC} software v12.5 \citep{arn96}. 
We used models with one- or two-temperature thermal {\it VAPEC} components \citep{smi01}
and the {\it TBABS} multiplicative model to describe the effect of
extinction by interstellar (and circumstellar) material (as
measured by the hydrogen column density $N_{H}$).
The plasma abundances for the {\it VAPEC} components
were fixed at the values adopted by the XEST study \citep{gud07}
to be typical for pre-main sequence stars\footnote{The adopted abundances,
relative to the solar photospheric
abundances given by \citet{and89}, are:
C = 0.45, N = 0.788, O = 0.426, Ne = 0.832, Mg = 0.263, Al = 0.5, Si = 0.309, S =
0.417, Ar = 0.55, Ca = 0.195, Fe = 0.195, Ni = 0.195.}.
For the extinction we used the standard interstellar
abundances in the {\it TBABS} model as listed in \citet{wil00}. 
In order to evaluate the goodness of our fits we choose to apply the 
C-statistic (a maximum likelihood method; \citet{cas79,wac79}) which is
better suited than the classic $\chi^{2}$ statistic for
low-count data. 

For two sources, \# 97 and \# 230, a two-temperature model
was required for an acceptable fit. The remaining spectra are well fit 
with a single thermal component. 
A few selected examples of the spectral fits are shown in Figure \ref{fig:spec}.
The spectral parameters are reported in Table \ref{tab:spec}. 
The hydrogen column densities
range from $\log \left(N_{H}\,[{\rm cm^{-2}}]\right) = 20$ to 
$\log \left(N_{H}\,[{\rm cm^{-2}}]\right) = 23.08$, corresponding to
visual absorptions between $A_{V} \la 0.1$~mag and $A_{V} \sim 65$~mag. 
These values are in good agreement with the estimates derived with \textit{XPHOT}.
The median
value is 22.04, corresponding to $A_{V} \sim 6$~mag. The plasma temperatures 
range from $\approx 0.5$~keV (6~MK) up to $\sim 15$~keV (170~MK). 
In Table \ref{tab:spec} we also report luminosities derived from the spectral 
fit, assuming a distance of 1.6~kpc. 
The intrinsic luminosities are calculated from the spectral fit parameters,
setting extinction to zero. The range of extinction-corrected 
intrinsic X-ray luminosities spans from 
$\log L_{t,c} = 30.56$~erg~s$^{-1}$ to 32.11~erg~s$^{-1}$ for the full
 [0.5-8.0]~keV band.

The X-ray properties can give us some clues about the nature of the sources.
The majority of the X-ray sources has plasma temperatures and X-ray luminosities
in the typical ranges found for YSOs in other star forming regions
\citep[see, e.g.,][]{pre05b}; together with the fact that most X-ray
sources have clear optical/infrared counterparts, this 
suggests that these X-ray sources actually are young stars in the S254-S258
complex.
However, the 
sources for which the spectral fit yields extremely high plasma temperatures
of $kT > 7.5$ keV deserve special attention.
Such very hard spectra are typical for extragalactic objects (AGNs),
but also sometimes found for very young stellar objects (protostars) \citep[see, e.g.,][]{ima01}.
As protostars are usually deeply embedded
in the clouds in which they formed, the location of these very hard
X-ray sources provides another hint towards their likely nature.
 
Considering these issues, source number 2 may well be an extragalactic contaminant.
First, the spectral fit yields an extremely high temperature,
but only moderate extinction. Second, this source has
 no optical/infrared counterpart, and, third, it is located at the periphery of the
S254-S258 region, well outside the boundaries of the molecular clouds. 
Furthermore, its X-ray spectrum can also be well reproduced with a power-law model,
as typical for AGN X-ray sources.
Similar arguments apply to source 30.

The other X-ray sources with extremely high plasma temperatures are located
in dense clouds, have infrared counterparts, and are thus probably deeply embedded very young stellar
objects.

\begin{figure}
   \centering
\includegraphics[width=8.5cm]{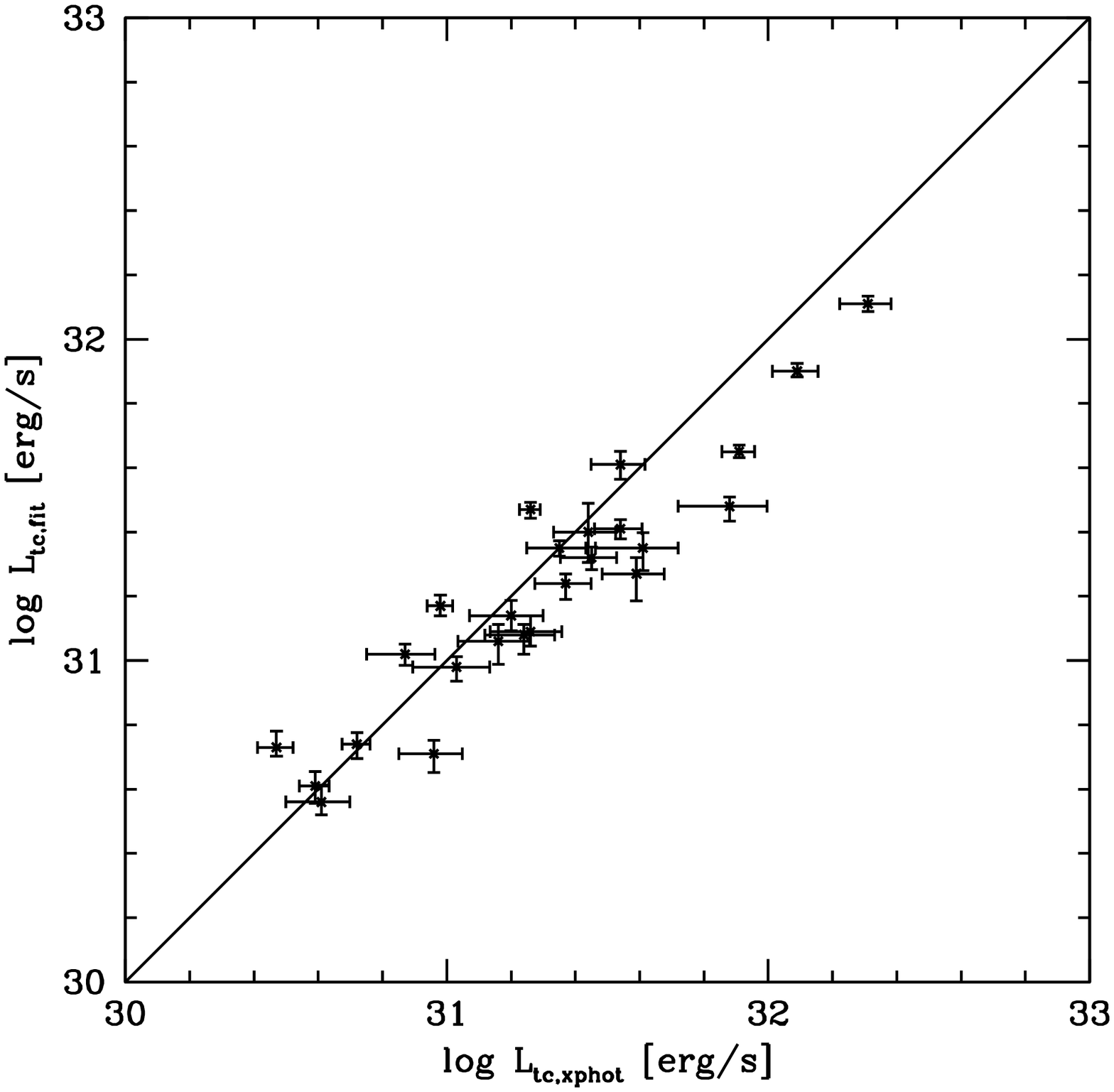}
   \caption{Comparison of the intrinsic full band [0.5-8.0] keV luminosities derived from the spectral fits
to the 25  bright sources (Table \ref{tab:spec}) to the intrinsic full band
   [0.5-8.0] keV luminosities determined with {\it XPHOT}.}
   \label{fig:cfrlum}%
    \end{figure}

It is interesting to compare the X-ray luminosities from the spectral fits
to those derived with {\it XPHOT}.
Figure \ref{fig:cfrlum} shows that 
the results from these two different methods agree quite well for the
majority of cases. Only for four of the highest luminosity objects
we find that {\it XPHOT} seems to systematically over-estimate the
X-ray luminosities by $\approx 0.1 - 0.2$~dex.
The generally good agreement
suggests that the X-ray luminosities derived with {\it XPHOT} are
reliable. We therefore will use the {\it XPHOT} results for our
analysis of the X-ray luminosity function presented below.

\begin{table*}
\caption{Spectral parameters of brighter sources: the spectral fit was performed with an absorbed thermal plasma model with one ({\sc tbabs*vapec}) or two components ({\sc
tbabs*(vapec+vapec)}).}             
\label{tab:spec}      
\centering          
\scriptsize
\begin{tabular}{rll r r@{$\pm$}l r@{$\pm$}l r@{$\pm$}l  ccccc}    
\hline\hline       
Source& CP / & CXOU J &\hspace{-3mm}Net Counts& \multicolumn{2}{c}{log $N_{H}$} & \multicolumn{2}{c}{$kT_{1}$} & \multicolumn{2}{c}{$kT_{2}$} & log $L_{s}$ & log $L_{h}$ & log $L_{h,c}$ & log $L_{t}$ &  log $L_{t,c}$\\ 
No. & IR class &	&	(counts)&\multicolumn{2}{c}{(cm$^{-2}$)}& \multicolumn{2}{c}{(keV)} & \multicolumn{2}{c}{(keV)}&(erg s$^{-1}$)&(erg s$^{-1}$)&(erg s$^{-1}$)&(erg s$^{-1}$) &(erg s$^{-1}$)\\
 (1)& (2) & (3) & (4) & \multicolumn{2}{c}{(5)} &\multicolumn{2}{c}{ (6)} & \multicolumn{2}{c}{(7)} & (8) & (9) & (10) & (11) & (12)\\
\hline                    
1   & yes &  061147.81+180312.6& 443.3  &21.81 & 0.15&     0.7 & 0.1    &\multicolumn{2}{c}{--}		       &30.89	&    30.14&	  30.19&       30.97&	    31.47  \\
2   & no &  061154.83+180016.5& 192.1  &22.04 & 0.46&    15.0 & 5.1    &\multicolumn{2}{c}{--}		       &30.26	&    31.08&	  31.11&       31.14&	    31.27  \\
3   & yes&  061159.55+175802.9& 159.6  &22.38 & 0.64&     2.6 & 1.2    &\multicolumn{2}{c}{--}		       &30.21	&    30.96&	  31.05&       31.03&	    31.40  \\
5   & yes / III&  061206.65+180336.6& 239.0  &21.41 & 0.16&     0.5 & 0.1    &\multicolumn{2}{c}{--}		       &30.88	&    29.55&	  29.57&       30.90&	    31.17  \\
27  & yes &  061230.43+175506.8&  81.8  &20.48 & 0.19&     9.4 & 9.9    &\multicolumn{2}{c}{--}		       &30.23	&    30.53&	  30.53&       30.70&	    30.71  \\ 
30  & yes &  061230.98+180336.9& 386.2  &22.07 & 0.13&     \multicolumn{2}{c}{15.0$^{*}$} &\multicolumn{2}{c}{--}    &30.62	&    31.46&	  31.49&       31.52&	    31.65  \\
32  & yes &  061231.17+180853.8 F&  99.5  &21.96 & 0.41&     3.8 & 2.3    &\multicolumn{2}{c}{--}		       &30.28	&    30.78&	  30.81&       30.90&	    31.09  \\
97  & yes / III&  061244.17+175914.0& 174.6  &21.56 & 0.26&     0.6 & 0.2    & 3.4 & 1.6       			       &30.57	&    30.45&	  30.46&       30.81&	    31.02  \\
102 & yes &  061244.76+175946.8& 156.6  &21.61 & 0.15&     3.1 & 0.8    &\multicolumn{2}{c}{--}		       &30.41	&    30.65&	  30.67&       30.85&	    30.98  \\
104 & yes &  061245.23+175810.3 F& 116.6  &\multicolumn{2}{c}{20.0$^{*}$}&    1.5 & 0.2    &\multicolumn{2}{c}{--}		       &30.45	&    30.06&	  30.06&       30.60&	    30.61  \\ 
113 & yes / III&  061246.71+175418.2&  86.6  &21.53 & 0.23&     0.7 & 0.1    &\multicolumn{2}{c}{--}		       &30.38	&    29.47&	  29.49&       30.43&	    30.73  \\ 
147 & yes / III&  061251.04+180644.1& 141.8  &20.00 & 16.03&    1.0 & 0.1    &\multicolumn{2}{c}{--}		       &30.69	&    29.66&	  29.66&       30.73&	    30.74  \\ 
183 & yes &  061253.48+175633.9&  91.2  &23.08 & 3.43&     54.2& 724    &\multicolumn{2}{c}{--}		       &28.43	&    31.13&	  31.35&       31.13&	    31.48  \\
190 & yes &  061253.73+175724.7& 373.2  &22.99 & 1.25&     5.4 & 1.2    &\multicolumn{2}{c}{--}		       &29.51	&    31.64&	  31.88&       31.65&	    32.11  \\
209 & yes &  061254.33+175927.7 F&  86.5  &22.86 & 1.99&     8.6 & 10.8   &\multicolumn{2}{c}{--}		       &29.13	&    30.99&	  31.16&       31.00&	    31.35  \\	
230 & yes &  061255.04+175930.7 F& 250.5  &22.30 & 0.53&     0.9 & 0.4    & 8.7 & 13.0       			       &30.44	&    31.11&	  31.18&       31.20&	    31.61  \\
255 & yes / I&  061258.21+175848.1&  91.7  &22.34 & 0.53&     2.6 & 0.9    &\multicolumn{2}{c}{--}		       &30.00	&    30.70&	  30.79&       30.78&	    31.14  \\
333 & yes &  061312.58+180706.2& 379.7  &22.56 & 0.46&     7.9 & 3.5    &\multicolumn{2}{c}{--}		       &30.33	&    31.60&	  31.70&       31.63&	    31.90  \\  
337 & no &  061316.04+175416.7& 139.9  &22.22 & 0.42&     3.9 & 1.6    &\multicolumn{2}{c}{--}		       &30.19	&    30.91&	  30.96&       30.98&	    31.24  \\
341 & yes &  061317.11+175344.9 F& 265.6  &21.74 & 0.17&     5.0 & 1.6    &\multicolumn{2}{c}{--}		       &30.70	&    31.16&	  31.17&       31.29&	    31.41  \\
348 & yes / II&  061325.57+175230.5& 163.1  &22.18 & 0.39&     1.6 & 0.3    &\multicolumn{2}{c}{--}		       &30.44	&    30.73&	  30.81&       30.91&	    31.35  \\ 
349 & yes / III&  061325.94+175923.5 F&  95.6  &\multicolumn{2}{c}{20.0$^{*}$}&    1.0 & 0.3    &\multicolumn{2}{c}{--}		       &30.48	&    29.73&	  29.73&       30.55&	    30.56  \\ 
354 & yes &  061327.41+175554.3& 125.2  &21.91 & 0.3&      5.6 & 4.1    &\multicolumn{2}{c}{--}		       &30.25	&    30.82&	  30.85&       30.93&	    31.08  \\ 
355 & yes &  061327.61+175517.8& 182.8  &22.11 & 0.35&     3.8 & 1.4    &\multicolumn{2}{c}{--}		       &30.38	&    31.00&	  31.05&       31.10&	    31.32  \\  
359 & yes &  061328.37+175604.4&  97.7  &22.15 & 0.52&     4.3 & 2.6    &\multicolumn{2}{c}{--}		       &30.07	&    30.75&	  30.80&       30.84&	    31.06  \\ 
\hline                  		  		 
\multicolumn{15}{l}{$^{*}$ indicates frozen parameters in the fit.}\\
\multicolumn{15}{l}{Col.\ (2):  presence of an optical or infrared counterpart, infrared class
from \textit{Spitzer} photometry (if available).}\\
\multicolumn{15}{l}{Col.\ (3): sources flagged with ``F'' showed flare-like variability during our 
observations; lightcurves are shown in Fig.~\ref{fig:lightcurves}.}\\
\multicolumn{15}{l}{Col.\ (4): absorbing hydrogen column density of the best-fit.}\\
\multicolumn{15}{l}{Cols.\ (5) and (6): plasma temperature(s) of the best-fit.}\\
\multicolumn{15}{l}{Cols.\ (7) to (11): X-ray luminosities (for an assumed distance of 1.6~kpc) 
in the soft (${s}$, [$0.5-2.0$] keV) band, the hard (${h}$ [$2.0-8.0$] keV) band,  
and the total (${t}$ [$0.5-8.0$] keV) band.}\\ 
\multicolumn{15}{l}{Absorption-corrected luminosities are denoted with the subscript ${c}$.}\\
\end{tabular}				  		  
\end{table*}

\subsection{X-ray source variability}

   \begin{figure*}
   \centering
   \includegraphics[width=7.0cm,height=3.8cm]{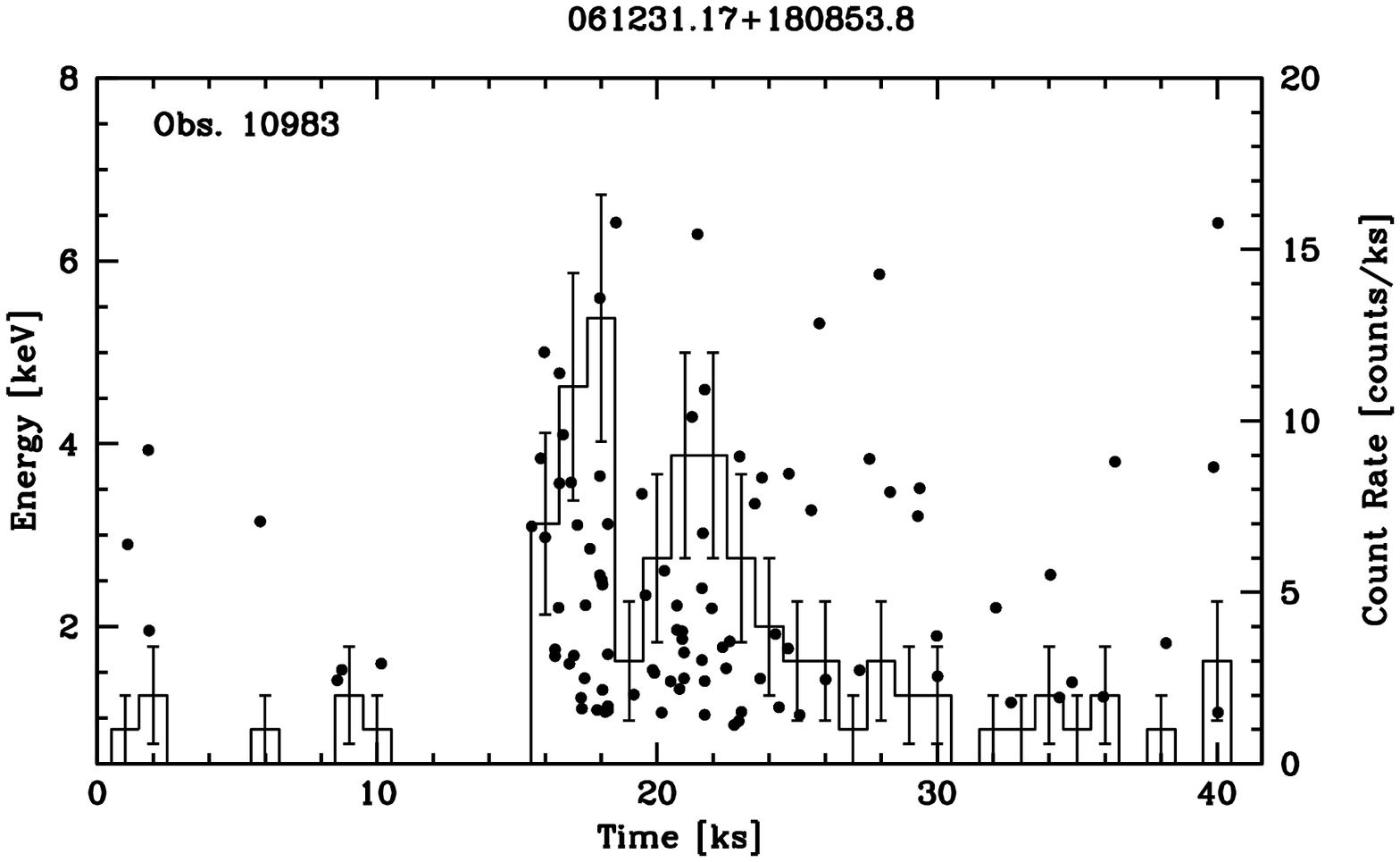} \hspace{2mm}
   \includegraphics[width=7.0cm,height=3.8cm]{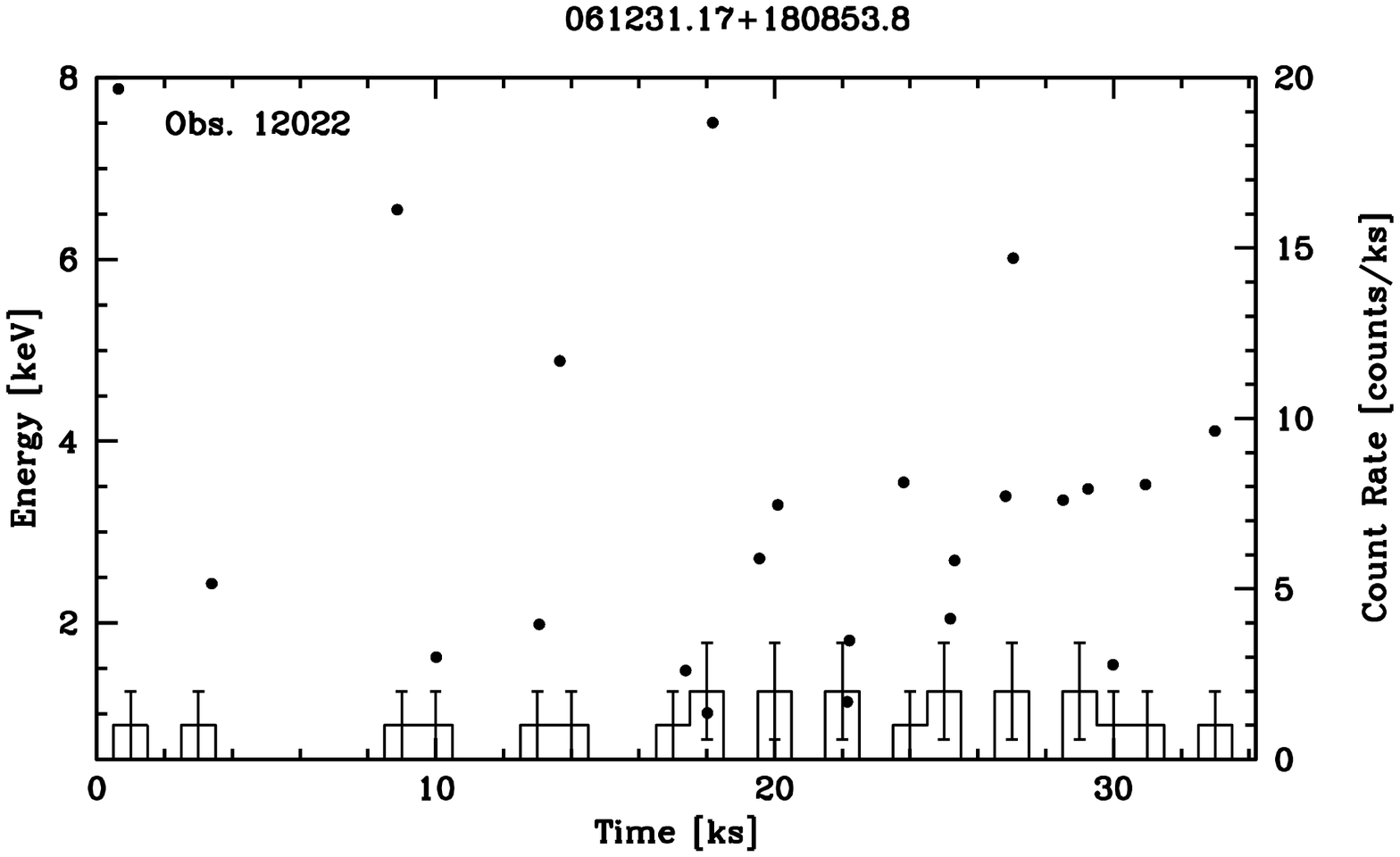} \\
   \includegraphics[width=7.0cm,height=3.8cm]{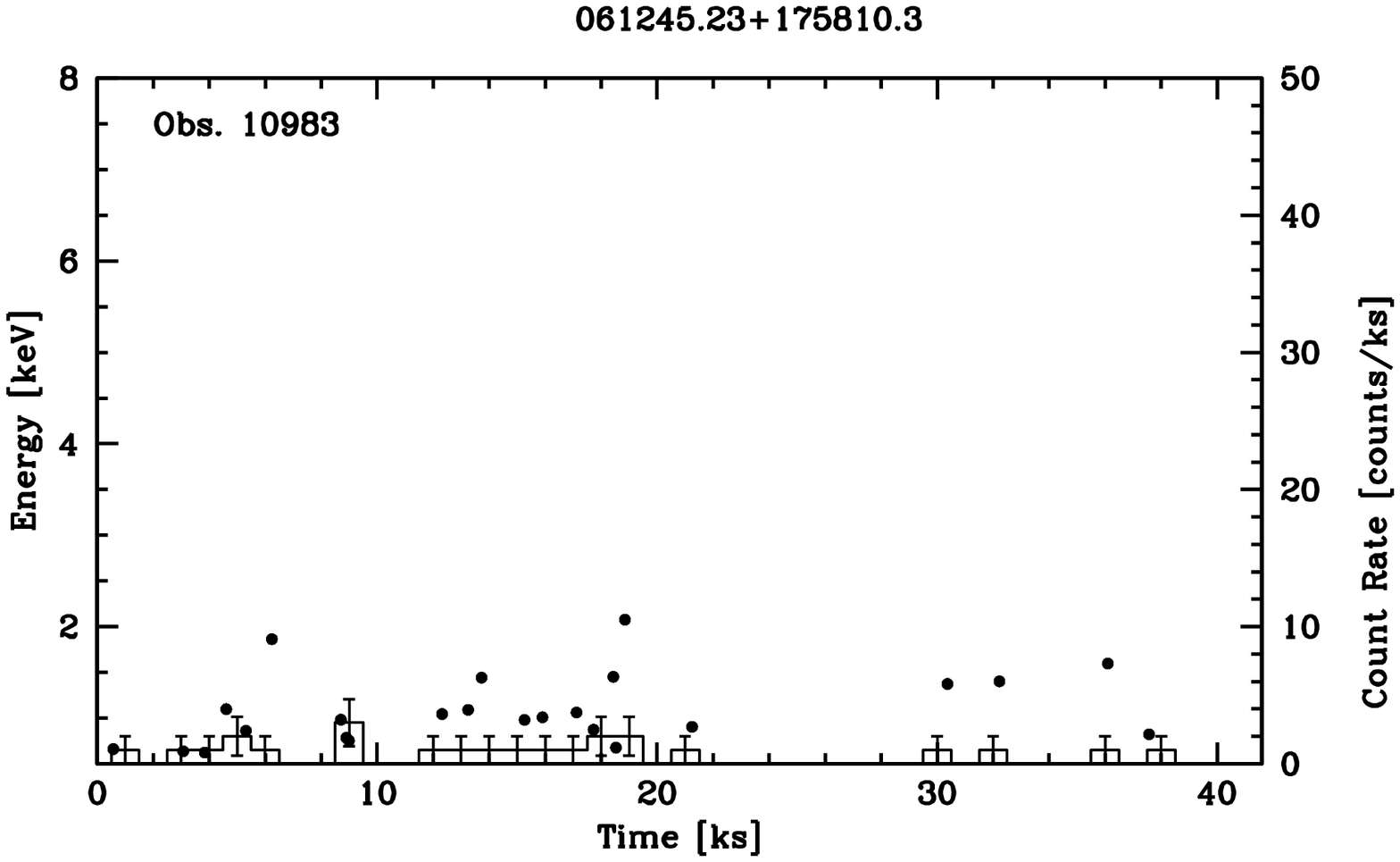}\hspace{2mm}
   \includegraphics[width=7.0cm,height=3.8cm]{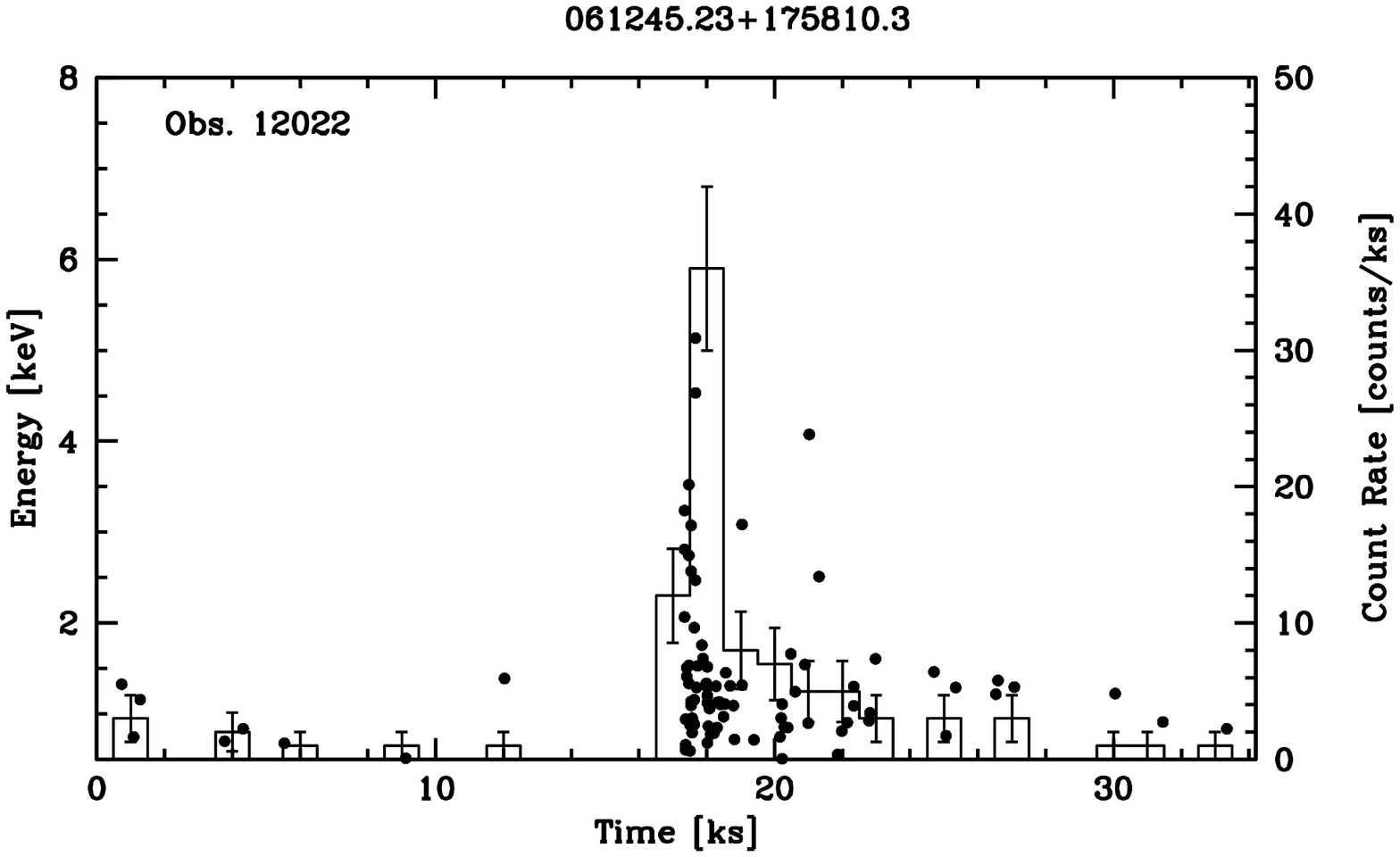}\\
   \includegraphics[width=7.0cm,height=3.8cm]{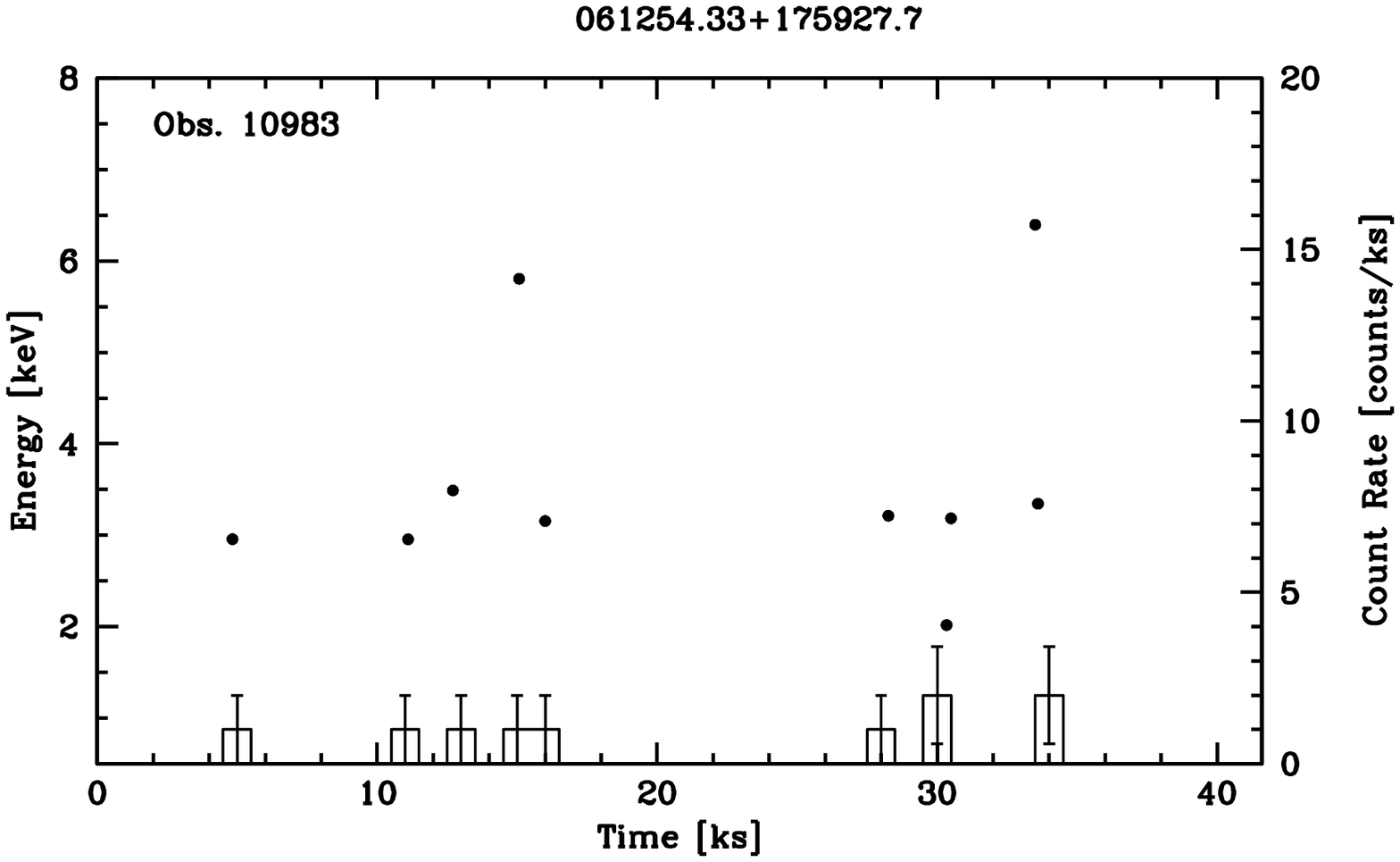} \hspace{2mm}
   \includegraphics[width=7.0cm,height=3.8cm]{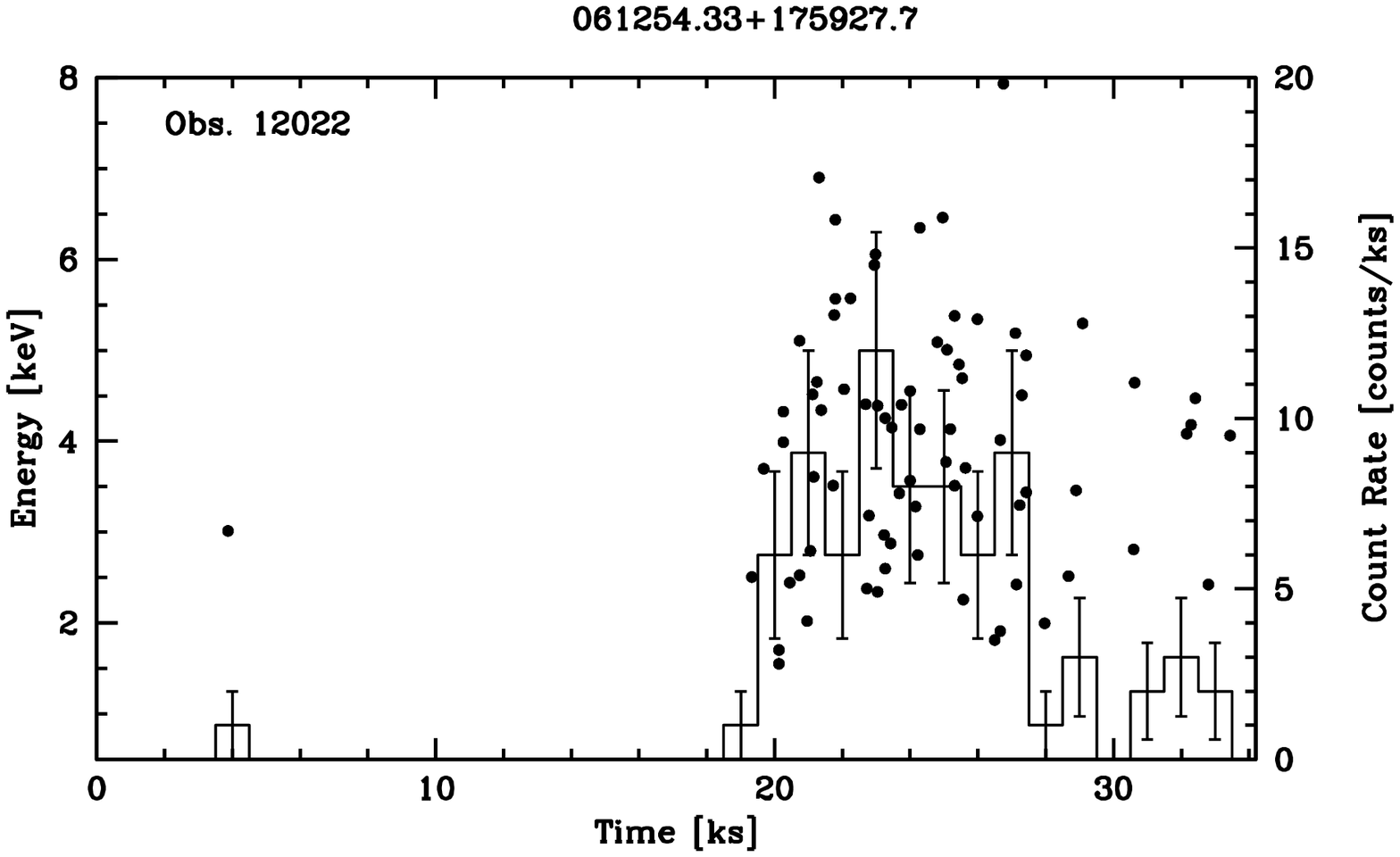} \\
   \includegraphics[width=7.0cm,height=3.8cm]{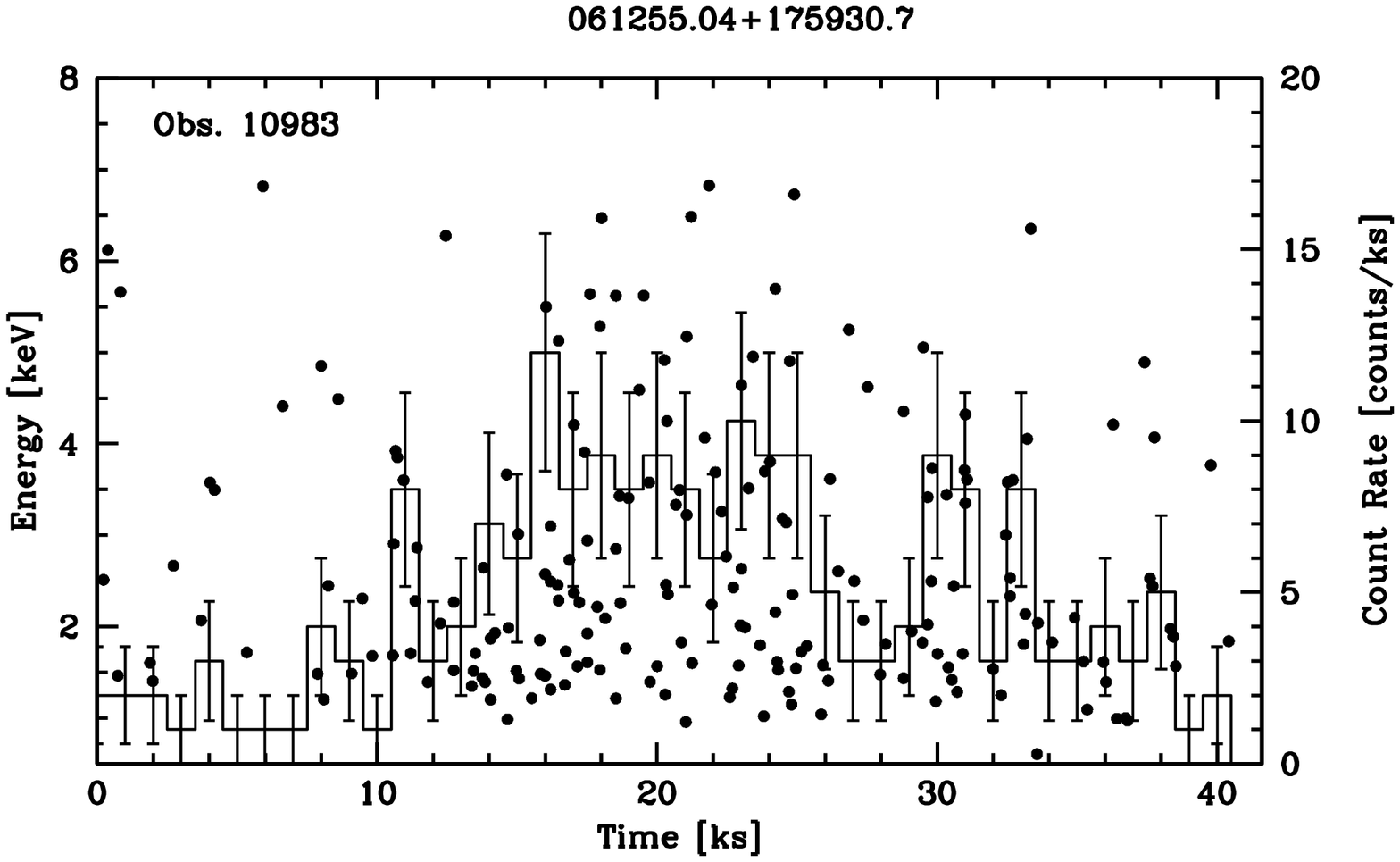}\hspace{2mm}
   \includegraphics[width=7.0cm,height=3.8cm]{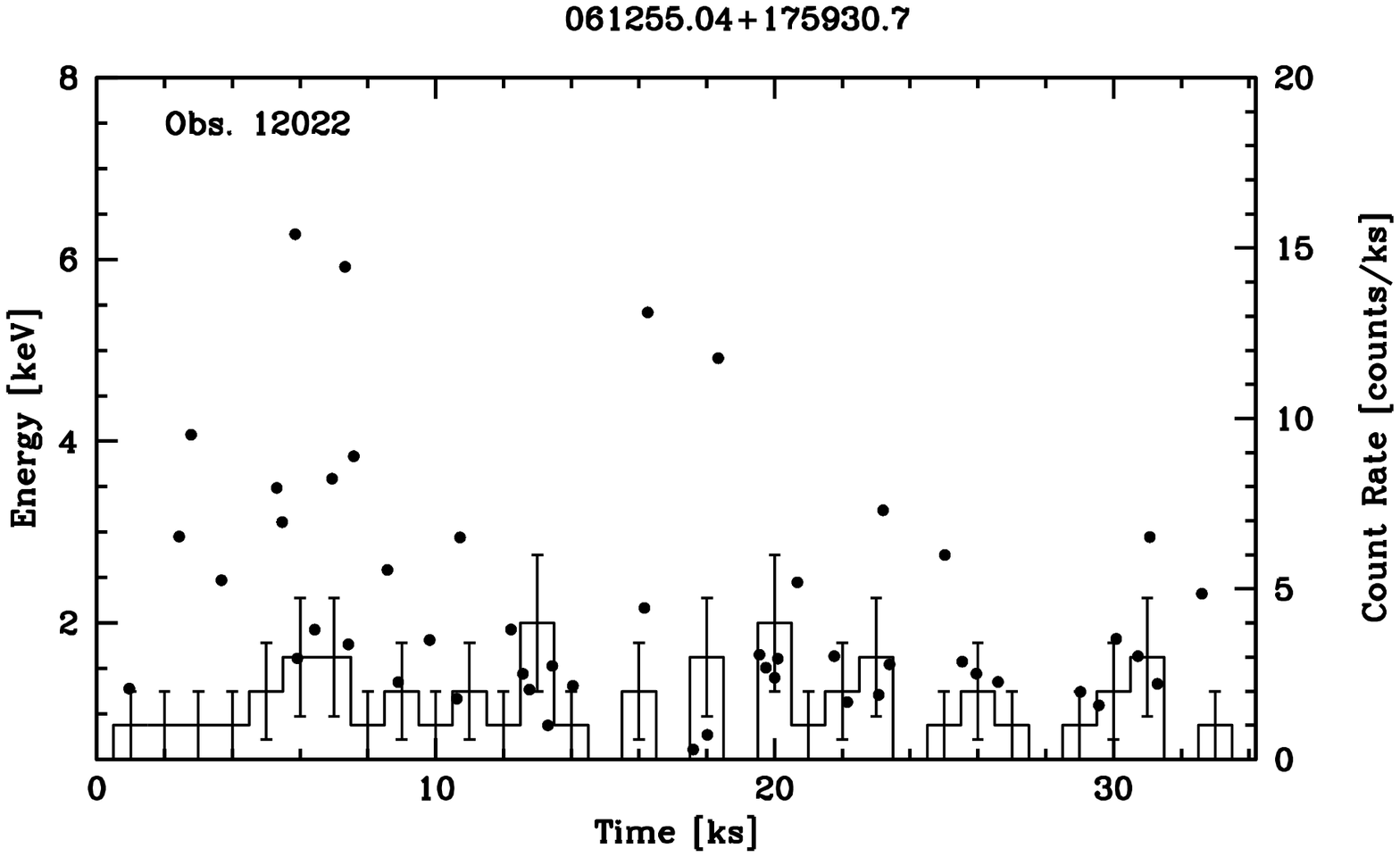}\\
   \includegraphics[width=7.0cm,height=3.8cm]{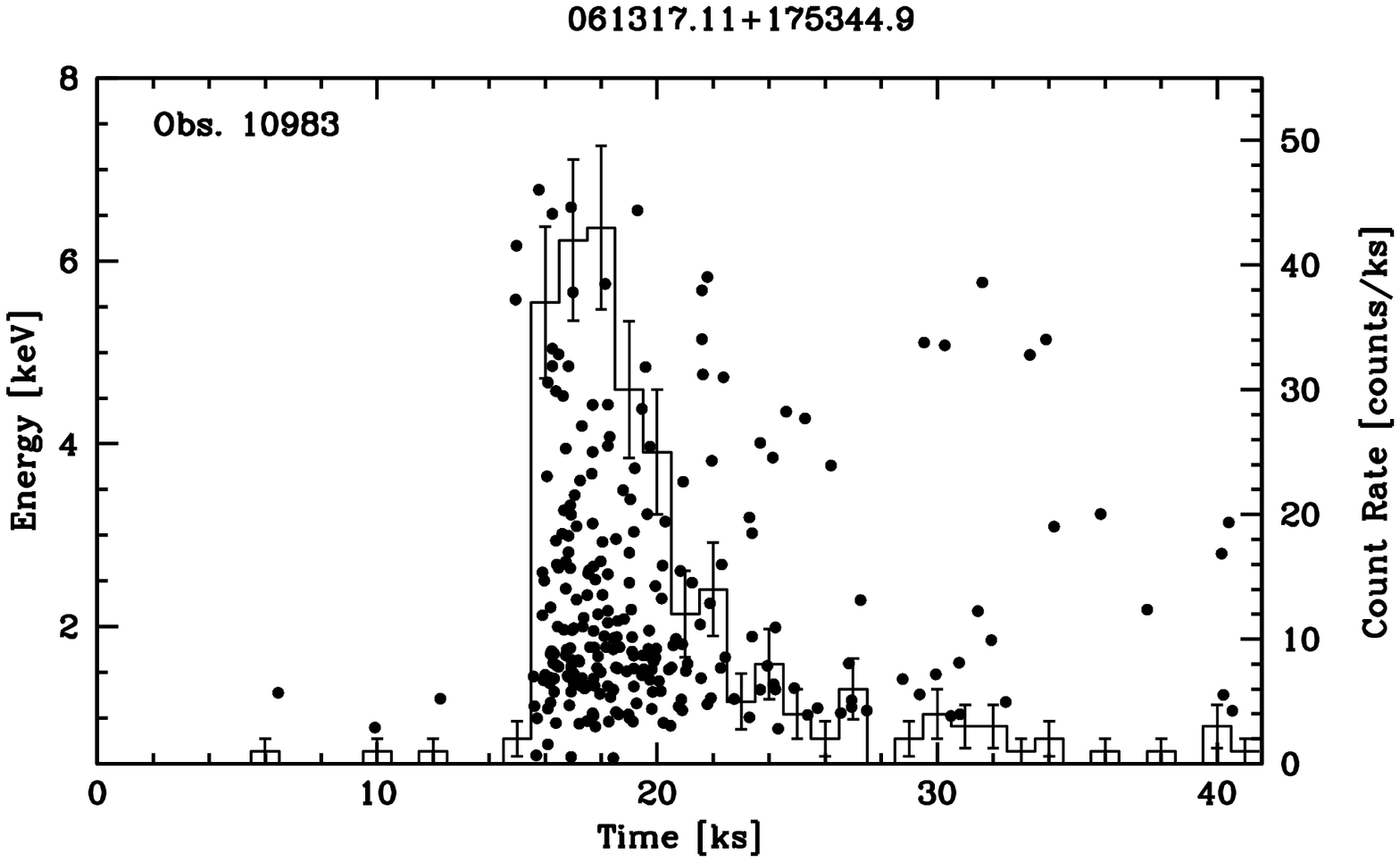} \hspace{2mm}
   \includegraphics[width=7.0cm,height=3.8cm]{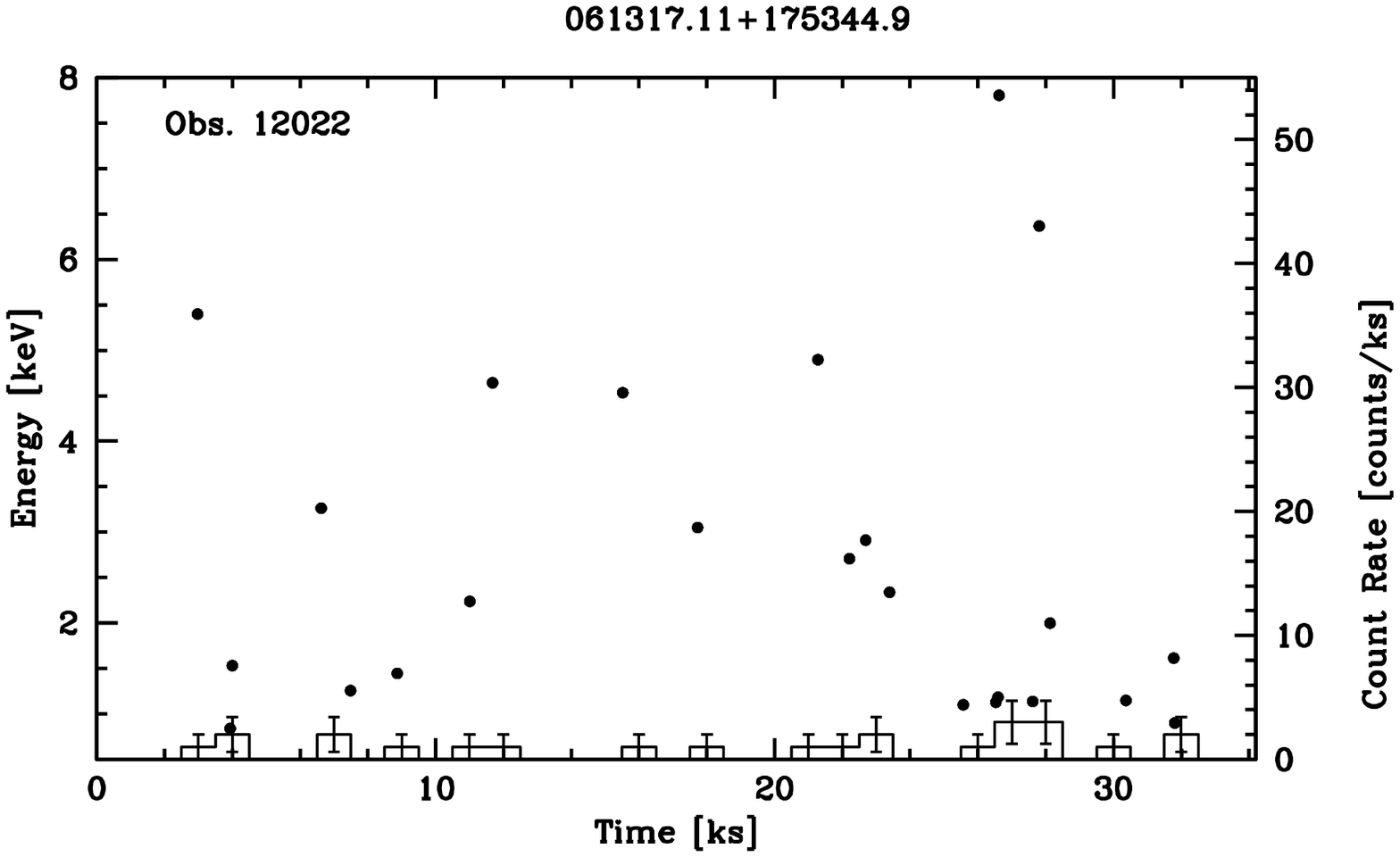} \\
   \includegraphics[width=7.0cm,height=3.8cm]{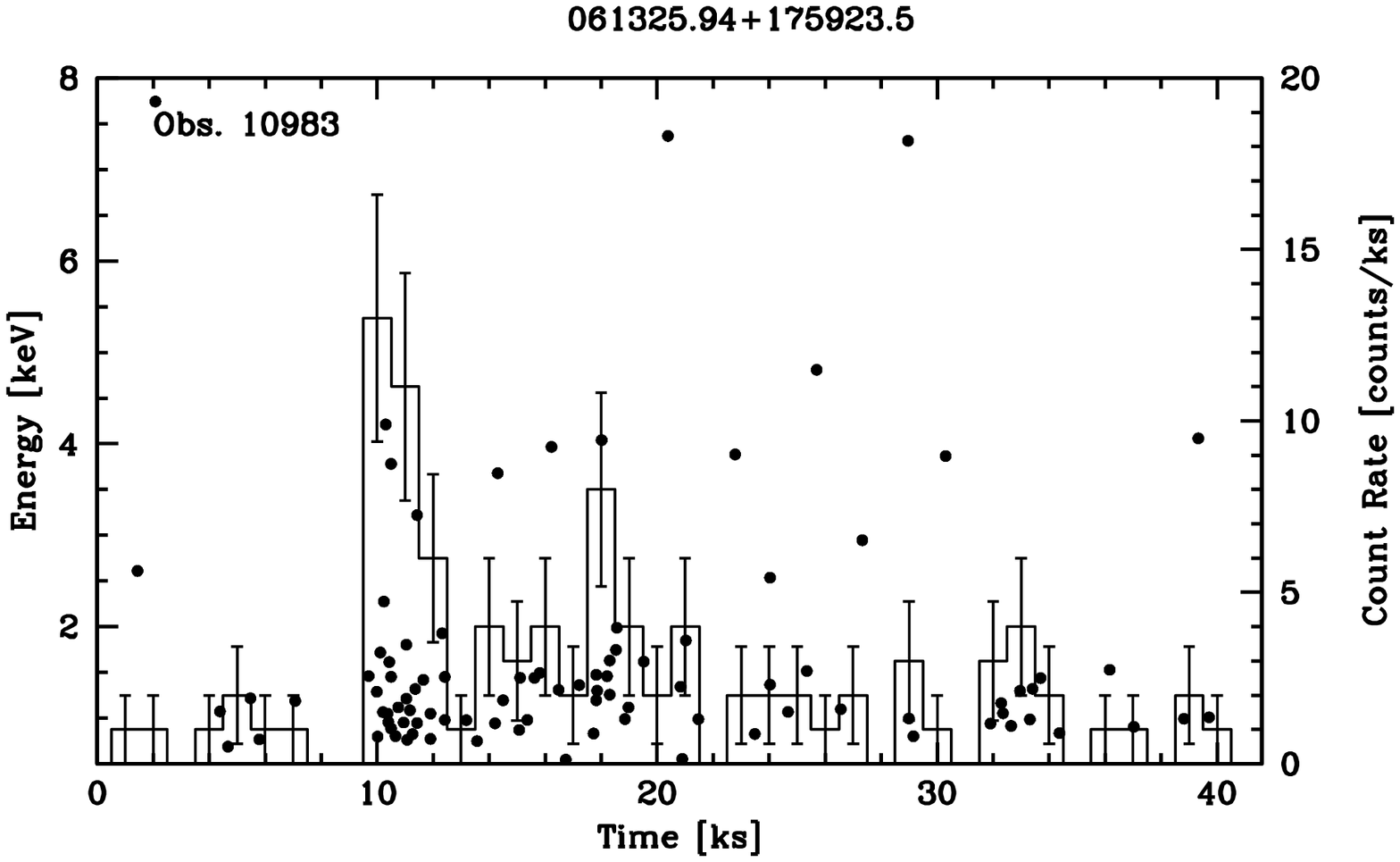}\hspace{2mm}
   \includegraphics[width=7.0cm,height=3.8cm]{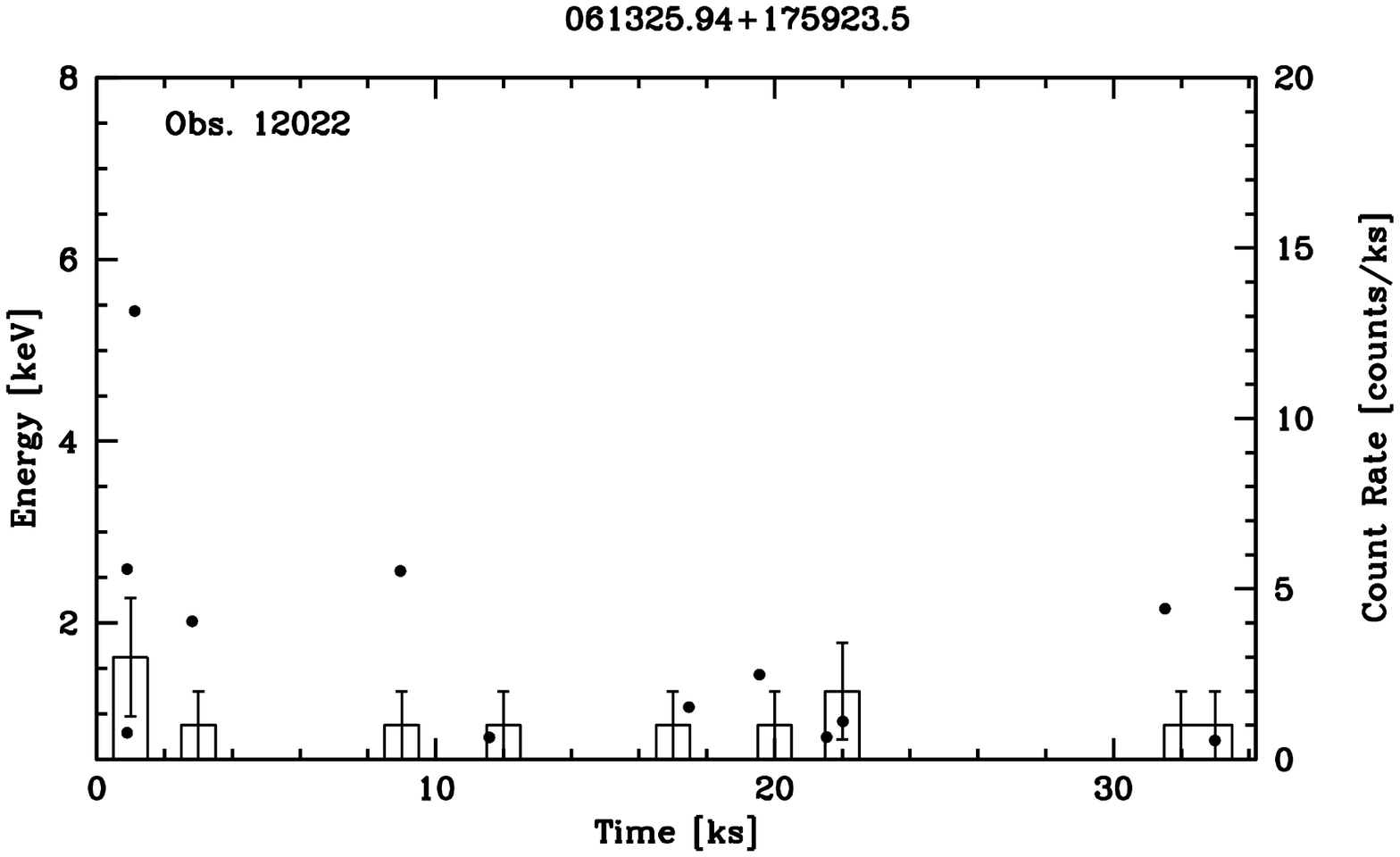}
   \caption{Lightcurves for six significantly variable sources. The solid dots
    show the arrival time (measured from the start of the observation)
    and the energy of each of the detected source photons. The histograms show
    the corresponding binned lightcurves.
   }
              \label{fig:lightcurves}%
    \end{figure*}

   \begin{figure}
   \centering
   \includegraphics[width=8.5cm]{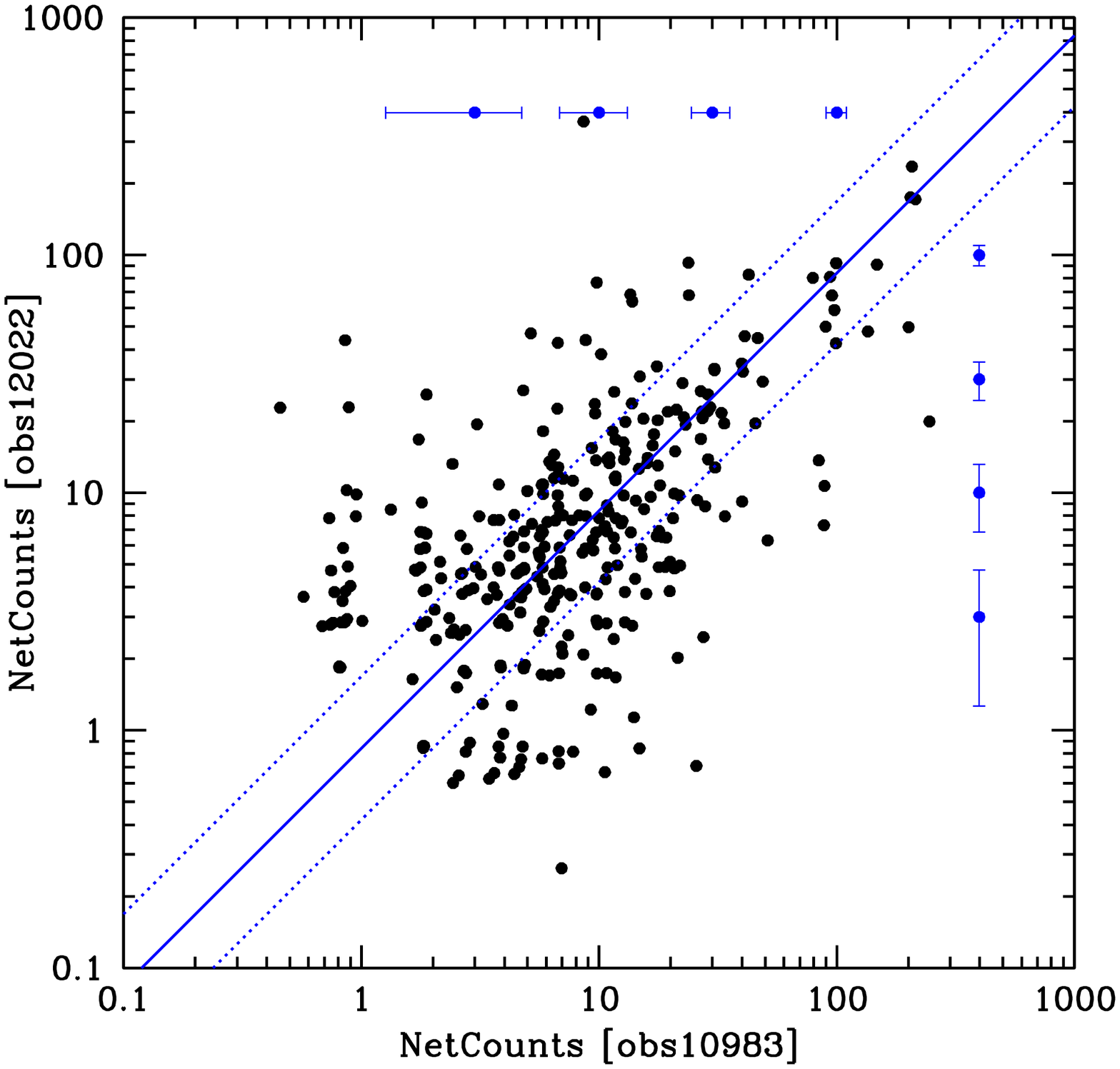}
   \caption{This plot shows for each X-ray source the net counts in the
first pointing compared to the net counts in the second pointing
(obtained about 4 days later). 
The sequence of blue error bars at the upper and right edge show
the Poisson statistical uncertainties for different numbers of
net counts. The solid line indicates the expected relation for
sources with constant count rates in the two pointings; the
dotted lines are offset by factors of 2.
   }
              \label{fig:variability}%
    \end{figure}

A first analysis of the time variability of individual X-ray sources
is performed by AE by comparing the arrival times of the individual
source photons in each extraction region to a model assuming
temporal uniform count rates.
The statistical significance for variability 
is quantified computing the 1-sided Kolmogorov-Smirnov
statistic (Col.~15 of Table~2). In our sample,
 23 sources show significant X-ray variability (probability of being constant
$P_{\rm const} <0.005$) and 19 are classified as possibly variable 
($0.005 <  P_{\rm const} < 0.05$).

The light curves of the variable X-ray sources show
a variety of temporal behavior; six of the most interesting lightcurves
are shown in Fig.~\ref{fig:lightcurves}. Five of these sources show
flare-like variability, i.e.~a fast increase of the count rate followed
by a slow exponential decay, as typical for
solar-like magnetic reconnection flares \citep[see, e.g.,][]{favata05,wolk05,p93}.
The other variable sources show
irregular variability or slowly increasing or decreasing countrates.
This kind of X-ray variability is typical
for young stellar objects
\citep[see, e.g.,][]{stassun06}.

We also investigated time variability by comparing the count rate for 
each source of the two single \textit{Chandra} pointings, i.e.~at a time
difference of about 4 days (see Figure \ref{fig:variability}). 
One can see that the majority of
sources show changes in the count rates by less than a factor of 2.
Only for 21 sources the count rates differ by more than 3$\sigma$ between
the two observations. The result is consistent with the
assumption that the X-ray emission from young stellar objects
is a superposition of many flares of different amplitude, where
weak flares are very frequent while very strong flares occur more
rarely, at rates
of about one such event per week \citep[e.g.,][]{get08}.


\section{Characteristics of the X-ray stellar population of S254-S258}\label{sec:pop}

\subsection{Optical and infrared counterparts of the X-ray sources}\label{sec:cc}

In order to identify counterparts of the X-ray sources in other wavelengths, we used the
optical images from the Digitized Sky Survey (DSS), the
Two Micron All Sky Survey (2MASS) point source
catalog, and the {\em Spitzer}-IRAC catalog from \citet{cha08}.
The results of the cross-correlation are reported in Table~5 (available in the electronic edition).
\onltab{5}{}
Our visual inspection of the red and blue DSS images gave
optical counterparts to 95 X-ray sources (i.e.~26\% of all 364 X-ray sources).
Our cross-correlation with the 2MASS Point Source Catalog lead to
231 near-infrared counterparts (i.e.~a counterpart fraction of 63\%).
Our cross-correlation with the
infrared catalog from \citet{cha08} yielded 293 infrared
counterparts, i.e.~80\% counterpart fraction.
For 58 X-ray sources  we did not find a counterpart in any of the
inspected optical and infrared images. 
Twelve of these sources are located in or very close to 
the central embedded cluster S255-IR. These X-ray sources may be
very deeply embedded protostars or young stellar objects located 
behind the dense molecular cloud clumps; the non-detection of optical/infrared counterparts
would then be related to very high extinction. 
The remaining 46 X-ray sources without known optical/infrared counterpart
outside this cluster show a rather homogeneous spatial distribution, as expected
for (mostly extragalactic) contaminants.

It is interesting to consider the infrared
classification of these sources based on the IRAC spectral energy distribution
(SED) slope determined by \citet{cha08}.
Unfortunately, the matching of our X-ray source-list
with this infrared catalog is not straightforward.
The catalog contains 26\,821 infrared sources. However, most of these are
only detected in the deep near-infrared images, and just
about 6400 of these are detected in the \textit{Spitzer} data.
Infrared classifications are only available for the
462 infrared sources that are detected in all four IRAC bands. 
The majority of the catalog entries are very faint NIR sources, and
many of these are probably background objects rather than young stars
in S254-S258. Furthermore, the
fact that only 4225 of the 26\,821 sources are detected in
all three of the $J$-, $H$-, and $K$-bands also suggest that there may well be a
significant number of spurious detections among the faint sources detected in only one
band.
This very high number of faint infrared sources produces
serious problems in any attempt to find the correct infrared
counterparts for our X-ray sources, since many
{\it Chandra} sources have more than one possible counterparts
within the X-ray error radius. In these cases, the closest
positional match is not necessarily the true counterpart.
Due to the increasing number of infrared sources at fainter
magnitudes, good positional
matches with very faint infrared sources may in fact                
be just chance superpositions of physically unrelated sources, 
and one of the other possible matches may be the true 
counterpart\footnote{We note that similar problems were 
encountered in an X-ray and infrared study
of the Carina Nebula; see \cite{Preibisch11} for a more 
detailed discussion.}.
A reliable identification of the infrared counterparts requires 
a sophisticated approach and will be addressed in the next step of our study.
Nevertheless, we can mention here the results of a preliminary
source matching, where we only considered the spatially closest match
to each X-ray source.
We find that 8 X-ray sources have closest matches classified
as Class I YSOs (embedded very young stellar objects
with infalling envelopes),
50 X-ray sources have closest matches classified
as  Class II YSOs (Classical T-Tauri stars, CTTs),
and 8 X-ray sources have closest matches classified
as Class III YSOs (``Weak line'' T-Tauri stars, WTTs) 
in the infrared catalog.

\subsection{The X-ray luminosity function}
\label{ssec:xlf}

  \begin{figure}
   \centering
\includegraphics[width=8.8cm]{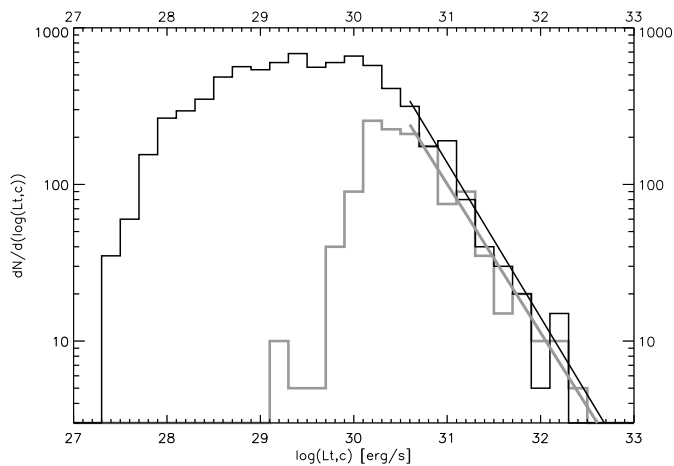}
 \caption{Comparison of the XLF of S254-S258 (thick grey line) to the XLF of the
Orion Nebula Cluster \citep[from the COUP data;][black line]{get05}. 
The straight lines show the results of the power-law fits to the distributions
in the luminosity range  $L_{\rm X} = 10^{30.5} \,-\, 10^{32.0}$~erg~s$^{-1}$.}
\label{fig:xlfcfr}%
    \end{figure}

The X-ray luminosity function (XLF) is the product of the
distribution of X-ray luminosities of stars with a given mass and
the number of stars per mass interval, i.e.~the initial mass function (IMF).
Although the correlation between stellar mass and X-ray luminosity shows
a considerable scatter \citep[see, e.g.,][]{pre05b},
X-ray studies of a large number
of young stellar clusters have shown that the XLF appears to be rather
universal  and constant in different environments
\citep[see][]{fei05b,get06,wan07}. 

To construct the XLF of S254-S258, we use
the intrinsic full band 
[$0.5-8.0$]~keV luminosities calculated by {\it XPHOT} (Table~3 column (7)). 
Our XLF of S254-S258 is shown in 
Figure \ref{fig:xlfcfr} and compared to the XLF for the stars in the 
Orion Nebula Cluster from the COUP \citep{get05}.
Obviously, the S254-S258 XLF peaks and turns over at a higher luminosity
(near $L_{\rm X} \approx 10^{30.3}$~erg~s$^{-1}$) than the COUP XLF,
because the X-ray detected sample for S254-S258 is incomplete for low-mass stars
due to the lower sensitivity as discussed above.
The slopes of the bright parts of these two distributions, however, can 
be seen to agree well.

For a more quantitative analysis, we performed power-law fits of the form
$d N / d(\log L_{\rm X}) \propto \alpha \times \log L_{\rm X}$
 for the observed distribution of luminosities.
We used the maximum-likelihood technique described by \citet{mk09},
that yields an estimate for the exponent from the observed distribution
function (i.e.~\emph{not} a fit of the histogram).
The resulting  power-law exponents for the distribution of X-ray luminosities
in the range
$L_{\rm X} = [ 10^{30.5} - 10^{32.0}]$~erg/s
are $\alpha = -0.95 \pm 0.09$ 
for the Orion COUP data, and $\alpha = -0.91 \pm 0.10$ for our 
S254-S258 data. This consistency confirms the results from comparisons 
of other regions \citep[e.g., the CCCP;][]{fei11}.

This result shows that it is reasonable to assume that 
the XLF of S254-S258 has a similar shape as the ONC XLF.
We can therefore make a quantitative estimate of the
size of the total young stellar population in the observed part
of S254-S258 by determining the vertical offset between the two distributions.
We find that the total population in the 
observed part of the S254-S258 complex is $\approx 0.7 \times$
of that in the ONC.
Since the total 
population of the Orion Nebula Cluster (within 2.06 pc, \citealt{hil98})
is about 2800 stars, the observed region\footnote{For comparison, we note that 
the diameter of the Orion Nebula Cluster ($\approx 30''$) would be $\approx 8'$ at the 
distance of 1.6~kpc.}
of the S254-S258 should contain
$\sim 2000$ stars in total.

\subsection{Spatial distribution of the X-ray sources}

   \begin{figure}
   \centering
\includegraphics[width=8.0cm]{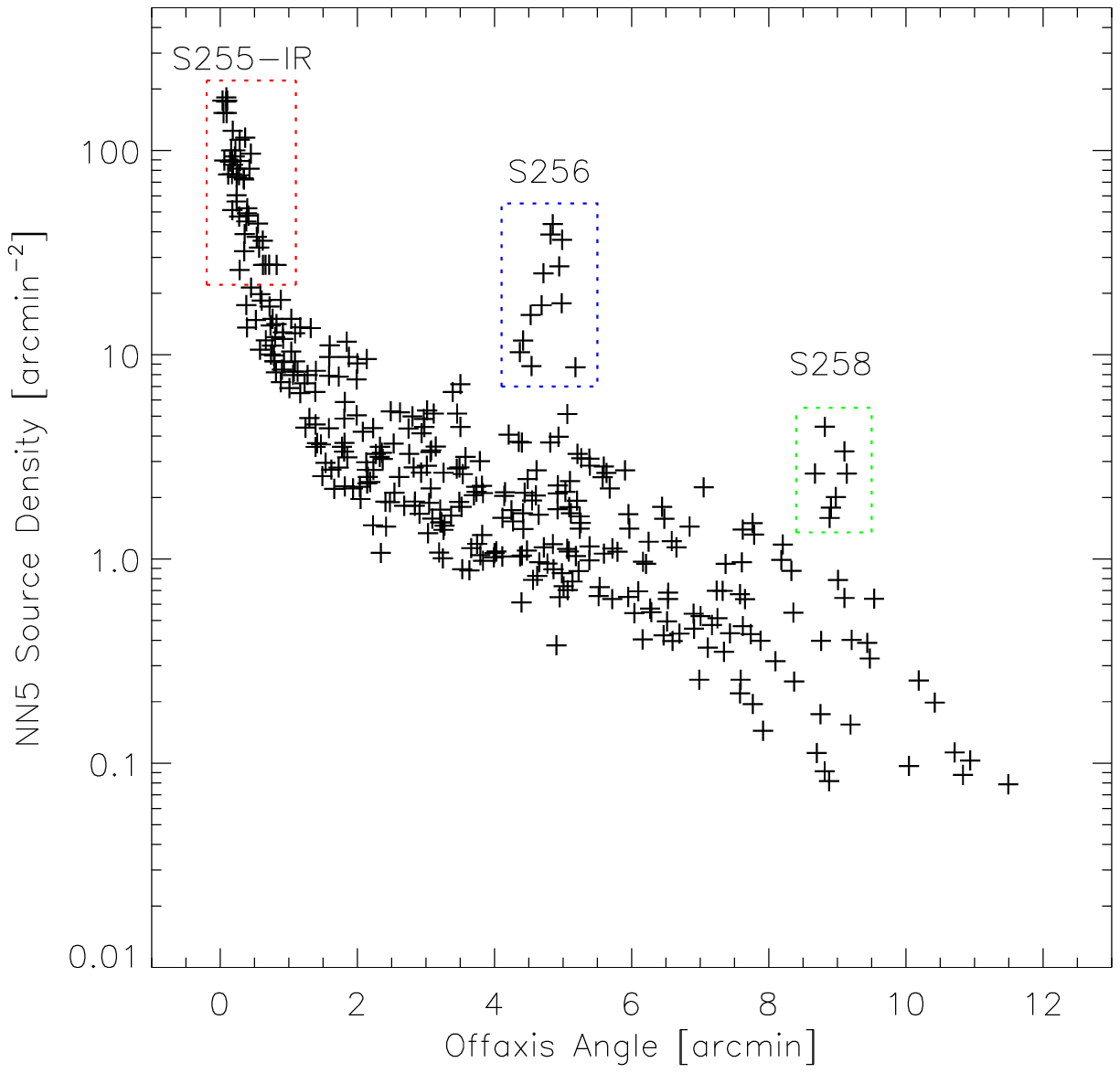}
\includegraphics[width=8.0cm]{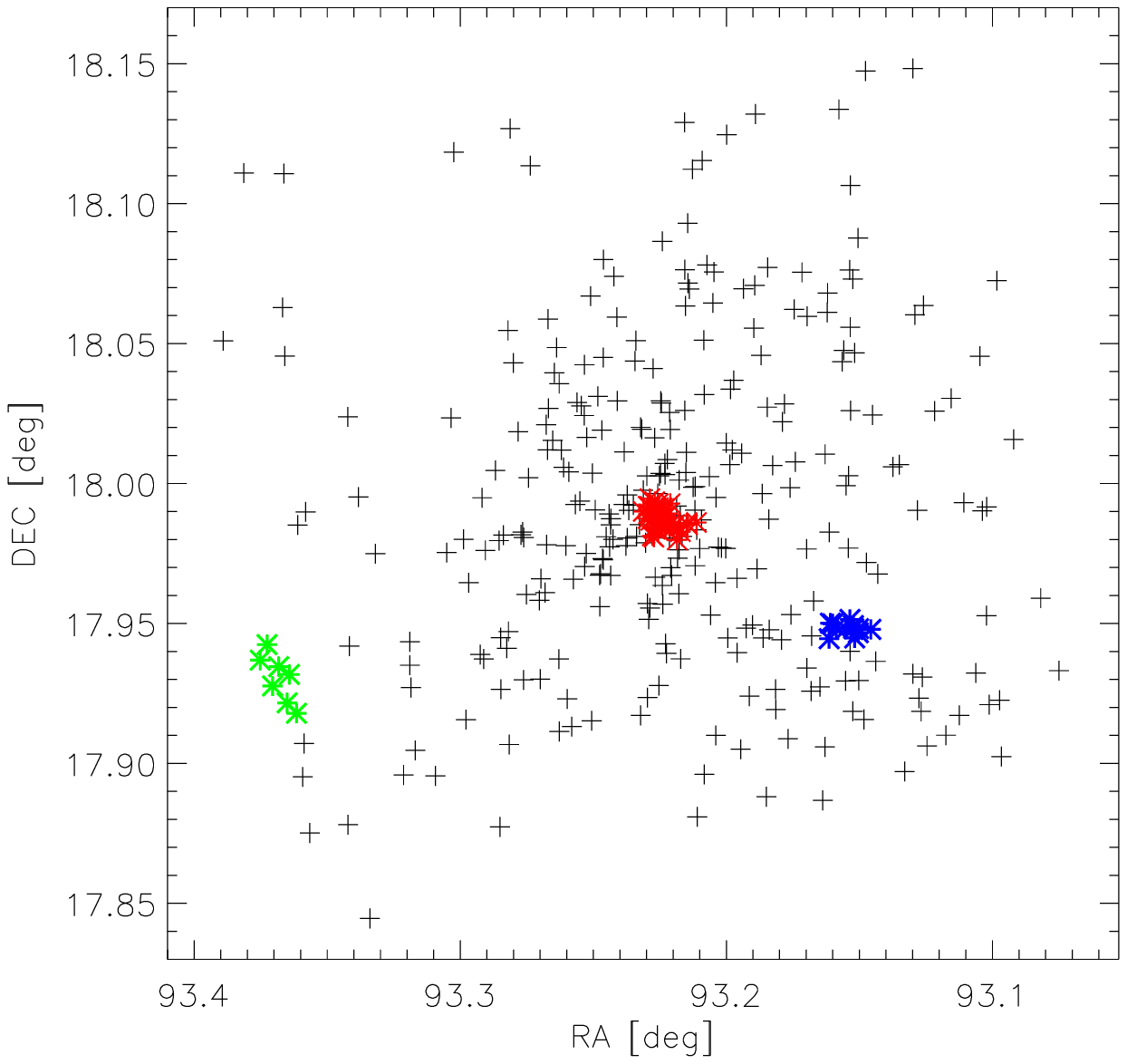}
   \caption{Top: Nearest neighbor analysis surface density at the location 
of each X-ray source plotted against the offaxis-angle. The general 
decrease of density with increasing offaxis-angle is related to
instrumental effects. The dot-lined boxes mark the members of the
three identified clusters. \newline
Bottom: Spatial distribution of the X-ray sources. The members of the three
clusters marked in the top plot are show by the colored asterisks.
	      } \label{fig:density}%
    \end{figure}

The spatial distribution of the 364 X-ray sources in the \textit{Chandra} field
shows a complex pattern.
Besides the prominent and dense central cluster S255-IR a few further
apparent clusterings as well as a widely distributed population
of X-ray sources that are spread homogeneously over the entire field-of-view of our
\textit{Chandra} observation can be seen.
For a quantitative characterization of the spatial distribution
we performed a nearest neighbor analysis \citep[see][]{CH85}
to identify statistically significant clusterings in an objective way.
The surface density estimator at the position of each source 
$i$ is
given by $\mu_j(i) = (j-1)\big/\left(\pi\,{D_j(i)}^2\right)\,,$
where $D_j(i)$ is the angular distance from
source $i$ to its $j$th nearest neighbor.
We used $j=5$ for our analysis; this value is large enough to
keep the fluctuations of the local density
estimates reasonably low  and at the same time allows to detect
groups with a minimum of $\sim 5$ stars.
For the interpretation of the resulting densities,
we have to take the spatial variations
of the detection sensitivity over the field-of-view
into account. 
The sensitivity is highest in the center, but
due to effects such as mirror vignetting and the
increasing width of the point-spread function, it
gets several times worse
near the edges of the ACIS field-of-view.
We therefore plot in Fig.~\ref{fig:density} 
the source density $\mu_5$ as a function of the
offaxis-angle.
The general trend of decreasing
source density with increasing offaxis-angle can be clearly seen.

Clusters can be defined as spatially confined groups of sources 
for which the local surface density clearly exceeds 
the values found at other locations in the image at similar
offaxis-angles.
The obvious dense central cluster S255-IR appears (as expected)
as a very prominent peak at low offaxis-angles
in Fig.~\ref{fig:density}.
Using a density threshold of $\mu_5 \ge 22\,\rm arcmin^{-1}$,
we find that 45 X-ray sources can be considered as
 members of this cluster.
Two further prominent peaks can be seen in the plot: 
one peak consisting of 12 sources near offaxis-angle 
$5'$ is caused by a cluster of sources
in the S256 region, while another peak around
offaxis-angle $9'$ with 7 sources represents
a clustering in the S258 region.

Our clustering analysis thus reveals three significant clusters,
that contain a total population of 64 X-ray sources.
The remaining 300 X-ray sources are thus in
a distributed, non-clustered spatial configuration.
As discussed above, up to $\approx 50$ X-ray sources may be unrelated
contaminants. This leads to a population of $\sim 250$
widely distributed X-ray detected young stars in the observed area.

\section{The size of the distributed X-ray population}

In order to check whether the distributed X-ray sources
may represent the low-mass stellar population associated
to the B0 stars in S255 and S257, we compare the size
of the distributed X-ray population with the expected number of
based on IMF extrapolations.
According to the \cite{kro01} field IMF model,
each B0 star 
\citep[$M \approx 15\,M_\odot$; see][]{martin05}
should be associated by $\approx 320$ low-mass stars 
$([0.1-2]\,M_\odot)$.
However, not all of these low-mass stars will be detected in our X-ray observation,
as the X-ray luminosities of young stars are related to the stellar mass.
The expected number of X-ray sources can be found
by comparing the X-ray detection limit to the typical X-ray luminosity
functions for young stars in specific mass ranges.

As discussed above, the X-ray detection limit of our data is
$L_{\rm X,min} \sim  10^{29.5}$~erg~s$^{-1}$ in the central part of the
observed field. However, this limit shows considerable systematic 
variations as a function of the  off-axis angle and
gets several times worse near the edges of the ACIS field.
If we consider the widely distributed population of X-ray sources, we
have to take into account that most of these sources are located
outside the central few arcmininute region of maximum sensitivity.
\citet{bro11} performed a detailed analysis of the spatial sensitivity variations
over the ACIS field. 
From the values for the completeness
limits in different off-axis angle slices in their Table 8 we find that
the area-weighted average of the completeness limit over the full field-of-view
is about 0.5~dex higher than the on-axis value.
This implies that the average X-ray completeness limit of our \textit{Chandra}
data for the widely distributed population is thus
$\log L_{\rm X} \sim 10^{30.0}$~erg~s$^{-1}$.

Since the X-ray luminosity functions for young stars are
very similar for most studied regions \citep[see][]{fei05b,get06}
we can assume that the young stars in S254-S258 follow the
same relations between stellar mass and X-ray luminosity
as established by the data from the 
\textit{Chandra} Orion Ultradeep Project \citep[see][]{pre05}.
This implies that we should detect $\approx 70\%$ of the young stars 
in the mass range $[0.5 - 2]\,M_\odot$
and $\approx 30\%$ of the stars in the mass 
range $[0.1 - 0.5]\,M_\odot$.

Since
the \cite{kro01} IMF predicts
$\approx 80$ stars with $[0.5 - 2]\,M_\odot$
and $\approx 250$ stars with $[0.1 - 0.5]\,M_\odot$
for each B0 star, the expected number of X-ray detected stars 
associated to the two B0 stars is
$ \sim 2 \times (0.7 \times 80 + 0.3 \times 250) = 262$.
The observed number of $\approx 250$ distributed X-ray sources (after correction for
background contamination) is actually quite close to this expectation value
and consistent with the assumption that these X-ray sources trace the low-mass
stellar population associated to the two B-stars.

\section{Conclusions}

Our deep \textit{Chandra} observation of the S254-S258 complex
led to the detection of 364 X-ray sources, about 50 of which are expected
to be background contaminants. 
The X-ray properties of most sources (luminosity, 
plasma temperature, and variability)
are in the typical ranges found for young stellar objects.
This supports the assumption that these
X-ray sources objects represent the population of young low-mass
stars in the S254-S258 complex.
Our analysis of the spatial distribution of the
X-ray sources with a nearest neighbor method
 reveals three significant 
clusters: the dense central cluster S255-IR,
and two smaller clusterings related to S256 and S258.
About 20\% of the X-ray sources are members of one of these clusters,
whereas the large majority $(\sim 80\%$) of the
X-ray sources traces a widely distributed population of young
stars.

The size of this distributed population of X-ray detected young stars
is in good agreement with the
expected X-ray source number based on the assumption that these stars trace 
the low-mass
population associated with the two early B-type stars in S255 and S257.
We would \textit{not} expect to see this distributed population in the context of the
models that two B0 stars have either formed in isolation or
were ejected from the central embedded cluster.
Our results therefore suggest that the two B-stars 
and the associated distributed low-mass stars represent a 
stellar population that is distinct from the embedded cluster of YSOs
in S255-IR.
This is in agreement with the model
scenario in which the observed star formation activity in the
dense embedded cluster located in the interaction zone between
the S255 and S257 \ion{H}{II} regions has been triggered
by the compression of the cloud due to the expansion of the
 \ion{H}{II} regions.

A detailed analysis of the optical and infrared
properties of the individual X-ray detected young stars
that can provide direct information on the
ages, masses, and the circumstellar disks around these
stars will be presented in a upcoming study.

\begin{acknowledgements}
This work
is based on observations obtained with the \textit{Chandra}
 X-ray Observatory, which is operated by the
Smithsonian Astrophysical Observatory for and on behalf of the National Aeronautics Space Administration (NASA)
under contract NAS8-03060. 
Our analysis presented in this paper was supported by the 
Munich Cluster of Excellence: ``Origin and Structure of the Universe''.
This publication makes use of data products from the Two Micron All Sky Survey, which
is a joint project of the University of Massachusetts and the Infrared Processing and Analysis Center/California
Institute of Technology, funded by the National Aeronautics and Space Administration and the National Science
Foundation, and of observations made with the Spitzer Space Telescope, which is operated by the Jet Propulsion
Laboratory, California Institute of Technology under a contract with NASA. 
\end{acknowledgements}

\bibliographystyle{aa}
\bibliography{17074}                                   

\end{document}